# Detection of Nanopores with the Scanning Ion Conductance Microscopy: A Simulation Study

Yinghua Qiu,[1,2,3]* Long Ma,[2,3] Zhe Liu,[2,3] Hongwen Zhang,[2,3] Bowen Ai,[2,3] and Xinman Tu[1*]

1. Key Laboratory of Jiangxi Province for Persistent Pollutants Control and Resources Recycle, Nanchang Hangkong University, Nanchang, Jiangxi, 330063, China

2. Key Laboratory of High Efficiency and Clean Mechanical Manufacture of Ministry of Education, National Demonstration Center for Experimental Mechanical Engineering Education, School of Mechanical Engineering, Shandong University, Jinan, 250061, China

3. Shenzhen Research Institute of Shandong University, Shenzhen, Guangdong, 518000, China

*Corresponding author: yinghua.qiu@sdu.edu.cn, tuxinman@126.com

# ABSTRACT

During the dielectric breakdown process of thin solid-state nanopores, the application of high voltages may cause the formation of multi-nanopores on one chip, which number and sizes are important for their applications. Here, simulations were conducted to mimic the investigation of *in situ* nanopore detection with scanning ion conductance microscopy (SICM). Results show that SICM can provide accurate nanopore location and relative pore size. Detection resolution is influenced by the dimensions of the applied probe and separation between the probe and membranes, which can be enhanced under large voltages or a concentration gradient.

## 1. Introduction

With the resistive-pulse technique, nanopores serve as accurate sensors for various objects, such as nanoparticles,[1] viruses,[2] vesicles,[3] and biomolecules.[4, 5] The mechanism originated from the Coulter counter principle which was first developed to detect blood cells in the 1950s.[6] When the objects pass through the nanopore, the ionic current shows a temporary decrease caused by the space occupation of the object inside the nanopore, which can be monitored by the patch clamp amplifier. Based on the resistive-pulse technique, label-free and high-throughput individual detection of analytes can be achieved. It has deserved considerable attraction since the potential application for DNA sequencing [7, 8] and protein sequencing.[9, 10]

Nanopores can be drilled by transmission electron microscopy (TEM), focus ion beam (FIB), or other kinds of high-energy beams.[11] These methods have relatively low yields and require professional equipment. With the developed dielectric breakdown (DEB) method, nanopores can be created conveniently with controlled sizes,[12, 13] which is of great importance for biosensing, nanofluidics, and related fields. For membranes thinner than 10 nm, the DEB method usually fabricates a nanopore under a weak working voltage. While, for the membrane with 30~50 nm in thickness, the applied voltage can reach 20 V or higher which can cause the formation of multi-nanopores on the membrane simultaneously due to potential material defects.[14] Due to the small pore diameter, the location and number of nanopores created by the DEB method usually cannot be easily determined by traditional nano-characterization methods, such as SEM or TEM, which can also cause damage or shrinking to the nanopores.[11, 15] To control the location of created nanopores, several location-controlled DEB methods have been developed, such as predrilling the membrane to form artificial defects,[16] adjusting the contact area with glass pipettes,[17] and forming local nanostructures with relatively weak materials.[18]

For the *in situ* detection of the location and number of the pores on the membrane, the fluorescence technique was tried by Meller et al,[19] and Yin et al.[20] With the help of high-resolution microscopy, pores with less than 10 nm can be found on the membrane with the increased strength of the fluorescence. However, this method has a low detection resolution for the pore location and size.[20] Based on the high resolution of ionic current, scanning ion conductance microscopy (SICM) was developed several decades ago to detect the surface properties in aqueous solutions based on the ionic current through the detecting probe.[21, 22] As a technique easy to operate, SICM has been used to detect micropores on membranes by Baker et al,[23, 24] and Venkatesh et al.[25] With the collected ionic current during the nanopipette scanning across porous membranes, track-etched pores with diameters from ~270 to ~930 nm were investigated and characterized.[23, 24] The detection resolution can be improved by applying a salt gradient across the porous membrane to increase the penetrated flux.[24]

Though SICM has a high space resolution,[26] the identification of nanopores with SICM is still challenging and rarely conducted. Here, 2D simulations were conducted to mimic the detection of the nanopore location and number on nanoporous membranes by the SICM, with a conical probe scanning above a porous surface containing one or two nanopores. Systematic simulations were conducted to explore the influence of the pore size, applied voltage, and number of pores on the detected probe current. Our results show that SICM can serve as a fast and easy *in situ* way for the nanopore detection which can give the exact locations and sizes of the nanopores.

## 2. Methodology

Fig. 1. Scheme of the simulation. Insets show the zoomed-in parts. A-Z respects the points on the boundaries of the system. Yellow regions show the glass pipette with 1 μm in length and 500 nm in base diameter. The tip diameter can be adjusted to 30, 60, and 100 nm.

The 2D simulation was conducted by solving coupled Poisson-Nernst-Planck (PNP) and Navier-Stokes (NS) equations with COMSOL Multiphysics package to model steady-state solutions of ionic current through the nanopipette and nanopores at room temperature (298 K) as described by Equations 1−4.[27-31]

$$\varepsilon \nabla^2 \varphi = -\rho_e = -F\sum_{i=1}^{N} z_i C_i \tag{1}$$

$$\nabla J_i = \nabla \left( -D_i \nabla C_i + \boldsymbol{u} C_i - \frac{F z_i C_i D_i}{RT} \nabla \varphi \right) = 0 \tag{2}$$

$$\mu \nabla^2 \boldsymbol{u} - \nabla p - \sum_{i=1}^{N} \left( z_i F C_i \right) \nabla \varphi = 0 \tag{3}$$

$$\nabla \boldsymbol{u} = 0 \tag{4}$$

where $\varepsilon$, $\nabla$, $\varphi$, $\rho_e$, $F$, and $N$ are the dielectric constant of solutions, gradient operator, electrical potential, volumetric charge density, Faraday constant, and number of ionic species. $z_i$, $C_i$, $J_i$, and $D_i$ are the valence, concentration, ionic flux, and diffusion coefficient of ionic species $i$ (including both cations and anions). $\mu$ and $p$ are the viscosity of solutions and pressure. $\boldsymbol{u}$, $R$, and $T$ are the velocity of the fluid, gas constant, and temperature, respectively.

Please note that we mainly focused on the characterization of DEB nanopores. Considering the smooth membrane surfaces, the chip membrane was simplified in our simulation model. Except for the nanopore, all other nanoscale structures were neglected on the membrane. Then, 2D simulations can be used to shed light on the nanopore detection with the SICM method.

Fig. 1 shows the scheme of the nanofluidic simulation system. All boundary conditions used in this work are listed in Table 1. For conical nanopipettes, the smaller and larger openings are defined as the tip and base, respectively. Table 2 lists the main dimensions of the simulation models. The length of the conical nanopipette was 1 μm, and its base diameter was 500 nm. Three tip diameters were applied 30, 60, and 100 nm.[32] The thickness of the pipette walls was 25 and 100 nm on the tip and base side respectively. −0.02 $C/m^2$ was selected as the surface charge density.[33] For the nanopore, 20 nm was selected for the length, and different diameters ranging from 2 to 50 nm were used. The surface charge density of the membrane was set as −0.005 $C/m^2$ to mimic the silicon nitride material.[33, 34]

All simulations were performed in KCl aqueous solutions assuming a dielectric constant of water of 80. Diffusion coefficients for potassium and chloride ions were assumed equal to the bulk value of 1.92 × $10^{-9}$ $m^2/s$ and 2.03 × $10^{-9}$ $m^2/s$, respectively.[35] The salt concentration is 0.1 M on the probe side. On the other side of the nanopore, the concentrations 0.1 M and 0.3 M were considered for the application of salt gradients across the membrane. The applied voltages of the electrodes are shown in Table 1.[23]

Following the mesh strategy used in our previous works,[29-31] a mesh size of 1 nm was used for the charged surfaces of the probe (GF, FL, KL, DE, EM, MN). 0.5 nm mesh was also used to confirm there is no dependence of the results on the mesh size (Fig. S1). For the charged boundary of the reservoirs (UV, WX, QR, ST) the mesh of 1.25 nm was chosen. For the inner surfaces of the nanopores, a finer mesh of 0.1 nm was selected to consider the ionic behaviors inside electric double layers near charged surfaces.

In our simulation cases, due to the combined influences of the short pipette length, weak surface charge density, and high salt concentration, no obvious ionic current rectification appears in the conical glass pipettes.[27, 36, 37] 0.3 V was selected as the probe voltage. The center reservoir was set as ground. A voltage scan from −0.5 V to 0.5 V was run across the nanopore, as shown in Fig. S2. We found that the detected current was much larger under negative voltages than under positive voltages. Then, −0.5 V was selected as the nanopore voltage. Please note that in our simulations, we haven't attempted to find some specific parameters to avoid the ionic current rectification through conical nanopipettes. The appearance of ionic current rectification in the nanopipette does not influence the following investigation of nanopore characteristics with the SICM due to the constant potential applied across the pipette.

**Table 1** Boundary conditions used in the numerical modeling. Coupled Poisson-Nernst-Planck and Navier-Stokes equations were solved with the COMSOL Multiphysics package.

| Surface | Poisson | Nernst-Planck | Navier-Stokes |
|---|---|---|---|
| AB | constant potential $\phi=0.3V$ | constant concentration $c_i= 100$ mM | constant pressure $p=0$ no viscous stress $\mathbf{n}\cdot[\mu(\nabla\mathbf{u}+(\nabla\mathbf{u})^T)]=0$ |
| AH,BC,CD,HG,JK, ON,UX,TZ | no charge $-\mathbf{n}\cdot(\varepsilon\nabla\phi)=0$ | no flux $\mathbf{n}\cdot\mathbf{N}_i=0$ | no slip |
| DE, GF, FL, KL, EM, MN | $-\mathbf{n}\cdot(\varepsilon\nabla\phi)=\sigma_w$ ($-0.02$ C/m$^2$) | no flux $\mathbf{n}\cdot\mathbf{N}_i=0$ | no slip $\mathbf{u}=\mathbf{0}$ |
| IJ,OP | constant potential $\phi=0$ | constant concentration $c_i=100$ mM | constant pressure $p=0$ no viscous stress $\mathbf{n}\cdot[\mu(\nabla\mathbf{u}+(\nabla\mathbf{u})^T)]=0$ |
| YZ | constant potential $\phi=-0.5\sim0.5V$ | constant concentration $c_i=100$ mM or 300 mM | constant pressure $p=0$ no viscous stress $\mathbf{n}\cdot[\mu(\nabla\mathbf{u}+(\nabla\mathbf{u})^T)]=0$ |
| UV, VW, WX, QR, RS, ST | $-\mathbf{n}\cdot(\varepsilon\nabla\phi)=\sigma_w$ ($-0.005$ C/m$^2$) | no flux $\mathbf{n}\cdot\mathbf{N}_i=0$ | no slip $\mathbf{u}=\mathbf{0}$ |

Table 2 Parameters used in the simulation models.

| Name | Value |
|---|---|
| Reservoir length | 5 μm |
| Pipette length | 1 μm |
| Base diameter of the pipette | 500 nm |
| Tip diameter of the pipette | 30, 60, 100 nm |
| Nanopore length | 20 nm |
| Nanopore diameter | 2~50 nm |

## 3. Results and discussion

Fig. 2. Simulated ionic current through nanopipettes. (a) Current-voltage (I-V) curves through the nanopipette located above a solid surface. The tip diameter of the probe was varied from 30 to 100 nm, and the base diameter was set to 500 nm. (b) Currents obtained with the nanopipette approaching a surface located in front of the probe at 0.3 V. (c) Currents obtained with the nanopipette approaching differently charged surfaces. In the inset, $D_{probe}$ denotes the inner diameter of the probe, which equals 60 nm. 0.1 M KCl was applied in all cases.

SICM has versatile applications in the characterization of surface properties, such as the morphology[38, 39] and charge density[40, 41] of solid surfaces as well as cell membranes.[42, 43] During the detections conducted by SICM, the collected data are the ionic current through the probe, which depends on the separation between the probe and the surface directly.[23, 24, 32, 44] Fig. 2a shows the current-voltage (I-V) curves through the glass pipette without a surface in front of the tip. Three tip diameters were used in the simulations from 30 to 100 nm. Due to the short length and large tip diameters of the nanopipettes, no obvious ionic current rectification appears through the conical nanopores.[27, 36, 37] Fig. 2b exhibits the current response during the approach of the probe to a charged silicon nitride membrane. With the separation between the probe and the surface decreasing, the current decreases due to the enhanced confinement which causes a higher access resistance.[33, 44] We also considered probes with different outer diameters of 260 nm and 110 nm, which have 100 and 25 nm in thickness at the tip, respectively. As shown in Fig. S3, with a larger outer diameter of 260 nm, the current decrease with the separation is more obvious because of the enhanced confinement by the thicker wall of the glass pipette.

Fig. 2c shows the dependence of the approaching current curves on the surface charge density of the membrane. When the probe-sample distance is larger than 5 nm the surface charge density of the membrane has little influence on the probe current. In our simulations, 0.1 M KCl solutions yield a Debye length of ~1 nm.[45] Under such spaces with low confinement between the probe and solid membrane, counterions inside electric double layers (EDLs) near the membrane surface cannot provide an effective contribution to the probe current.[29] As the probe-sample distance decreases to sub-5 nm, the surface charge density of the membrane starts to influence the probe current.[33] This is due to that with the confinement between the probe and membrane increasing, counterions inside EDLs near the membrane surface can have a considerable contribution to the total current.[29] In the case with the probe-sample distance of 0.5 nm, as the surface charge density of the membrane enhances from −0.01 $C/m^2$ to −0.05 $C/m^2$, the probe current has an increase of ~10% (Fig. S4).[33]

Fig. 3. Behaviors of ionic current through the glass nanopipette during its scanning over nanopores with different sizes. (a) Current detection by a probe scanning in the lateral direction of nanopores with different diameters at 10 nm above the surface. The lateral distance represents the separation between the probe axis and the pore axis. (b) Ionic current through the probe obtained at different probe-membrane separations. The probe approaches the membrane with a nanopore along the pore axis. (c) Relative current increases with the probe approaching the membrane with a nanopore compared with the current without a nanopore on the membrane. (d) Relative current increases with the nanopore size. The current increase ($\Delta I$) was the subtraction of the probe current above a nanopore by that obtained above the membrane without a nanopore. $\Delta I/I$ was obtained using $\Delta I$ divided by the probe current above a membrane without a nanopore. The inset shows the simulation scheme. $D_{pore}$ denotes the diameter of the nanopore.

For the nanopore detection by SICM with our 2D simulation models, the detection performance was considered in two directions, i.e. the axial and lateral directions. As shown in Fig. 3, five nanopores with different diameters varying from 2 to 30 nm were used to mimic nanopores on 20-nm-thick silicon nitride membranes created by the DEB method. Current detection was conducted by scanning the probe over the surface in the lateral direction of the nanopore. The glass nanopipette was 60 nm in the inner diameter of the tip. The separation between the probe and the surface was set at 10 nm. From the obtained current behaviors (Fig. 3a), when the probe is close to the pore, a much larger current is detected than that obtained when the probe is located far away from the pore. The peak value appears when the probe is above the pore with its axis in line with the pore axis. From our results, as the pore diameter changes from 2 to 30 nm, the full width at half maximum (FWHM) in the current trace varies from ~90 to ~110 nm, which is much larger than the pore size and the probe size. As shown in Fig. S5, the FWHM also depends on the separation between the probe and the membrane surface. To get a higher resolution, a smaller probe-sample separation is required. For the current trace obtained by the probe with a larger outer diameter, the FWHM is even larger (see below). Similar to the results in Fig. 2, the current depends on the probe-sample separation. At a probe-sample separation of 10 nm, even for the smallest pore with 2 nm in diameter, the current increase can achieve more than 20% compared with the case without a nanopore on the membrane (Fig. 3c). In this case, the pore location can be easily identified through the scanning of the whole porous membrane.

Also, the obtained probe current correlates to the nanopore size. During the probe scanning over a larger nanopore, a higher ionic current can be obtained, which provides a convenient *in situ* way to size the nanopore on the membrane. The increase of the ionic current through the probe results from the additional contribution of ions transported through the nanopore.[24, 46] With the nanopore size increase, more ions transported through the nanopore can balance the current decrease caused by the separation shrinking between the probe and the membrane. This is important because, after the DEB fabrication, the size of the nanopore cannot be identified just by its conductivity due to the unknown number of nanopores on the membrane.[47] While under an applied voltage, concentration polarization can form across the thin nanopore,[45] i.e. cations and anions get enriched and depleted at both pore ends (Fig. S6). At a small distance of the probe over the nanopore, the probe current has a close dependence on the surface charge density (Fig. S7). From Fig. S8, due to the induced stronger concentration polarization, the probe current becomes smaller when it is located above a strongly charged nanopore. Please note that in the practical detection, due to the unknown pore number on the membrane, direct nanopore sizing with ionic conductance is not feasible. Here with the SICM detection the collected probe current can only shed light on the relative pore size and the pore number (see below).

Referring to the previous work by Chen et al.,[23, 24, 48] during the practical detection many factors, such as pore roughness, and contamination, limited the detection sensitivity of the SICM. Only micropores with diameters larger than ~270 had been imaged with SICM. So, in our simulated cases, the nanopores with ultra-small diameters are challenged to be detected, though with the consideration of the smooth surface of the SiN membrane.

Fig. 4. Behaviors of ionic current through the glass nanopipette during its scanning over nanopores under a concentration gradient of 300: 100 mM KCl. (a) Current detection by a probe scanning in the lateral direction of a nanopore of different diameters with 10 nm above the surface. The lateral distance represents the separation between the probe axis and the pore axis. (b) Relative current increases as the probe approaches the membrane obtained with and without a concentration gradient across the nanopore. $\Delta I$ is the current increase by locating the probe over a nanopore. (c) Current increase ratio under a concentration gradient to that without a concentration gradient. (d) Scheme of simulation under a concentration gradient across the nanopore.

During scanning of the probe in the lateral direction, the increase of the probe current can be induced by the additional ions near the pore orifice transported through the nanopore. To enhance the current increase phenomenon, a concentration gradient was applied across the nanopore by setting 300 mM KCl on the membrane side without the pipette. Under such a concentration gradient of 300: 100 mM KCl across the membrane, many more counterions can be transported along the electric double layers near negatively charged pore walls[29] which results in a much larger ionic current through the probe. As shown in Fig. 4, the FWHM in the current trace varies from ~100 to ~110 nm, similar to that in the situation without a concentration gradient. From the comparison between both cases: 100: 100 mM KCl and 300: 100 mM KCl, the increase of probe current can be enhanced significantly, even for the situation with small nanopores of sub-10 nm in diameter. This phenomenon sheds light on an easy way to improve the detection resolution during the nanopore detection with the SICM technique, especially in the case where large voltages cannot be applied due to the potential expansion of the nanopore under high voltages.[49, 50]

Fig. 5. Simulated detection of double pores on the membrane. (a) Current traces obtained by scanning the probe over two 20 nm-in-diameter nanopores. Their axis separation was set to 50 nm and 100 nm. The lateral distance represents the separation between the probe axis and the center of two nanopores. (b) Detection of two pores with different sizes of 20 nm and 50 nm under different probe-membrane separations. Two nanopores have an axis separation of 100 nm. The separation between the glass probe and the membrane varies from 10 to 100 nm. The solid lines at the lower left show the probe current over the membrane without nanopores. (c) Detection of two nanopores with one 20 nm in diameter and the other with varied diameter from 5 to 50 nm at the probe-membrane separation of 10 nm. (d) Relative current difference between the nanopore with varied diameter from 5 to 50 nm and the 20 nm-diameter nanopore. The inset shows the illustration of the simulation model.

Detection of single nanopores with SICM can provide us with the accurate location of the pore on the membrane and its relative size. To explore the lateral detection of SICM, i.e. the ability to identify multi-pores on the membrane,[14] double-pore membranes were used underneath the probe. As shown in Fig. 5a, two nanopores with 20 nm in diameter were detected by scanning the probe of 60 nm in tip inner diameter over the membrane. From our results, if the two pores are located very near to each other, such as 50 nm, only one current peak appears in the probe current trace. In this case, SICM cannot identify the accurate location of each pore with the selected glass nanopipette. This is attributed to the large FWHM as shown in Fig. 3a. The large inner diameter of the probe limits the detection resolution. In the other case of both pores located 100 nm away from each other, the locations of individual nanopores can be clearly detected. Fig. 5b shows different scanning current traces over a membrane with two nanopores with different sizes of 20 and 50 nm, but different probe-sample distances. From the results, due to the larger current response caused by the larger pore, both individual nanopores can be identified.[23] While the identification can only be achieved with small separations between the probe and the membrane. Under those separations larger than 50 nm, the current increase profile becomes blurred near the pore boundary.

For cases with two nanopores on the membrane, we explored the nanopore detection under the influence of the diameter of its nearby nanopore. As shown by the inset of Fig. 5d, two nanopores are located on the membrane with an axis separation of 100 nm. The diameters of the left and right nanopores were denoted as $D_{pore1}$ and $D_{pore2}$. Here, $D_{pore1}$ was kept at 20 nm, and $D_{pore2}$ was varied from 5 to 50 nm. From Fig. 5c, during the probe scanning over both nanopores, the locations of both nanopores can be identified from the profile peaks. Also, the relative size of the nanopores can be evaluated from the current values. When the probe is located over a larger nanopore, a higher current is obtained. Through the comparison among current profiles, the detection of nanopore 1 has a weak dependence on the diameter of the nearby nanopore 2. From the probe current obtained over both nanopores, the relative current difference can be plotted with the nearby pore size, which was calculated by the peak current difference over the probe current above nanopore 1. Based on the monotonous dependence of the current difference on the pore size of nanopore 2, the quantitative relationship between the diameters of nearby nanopores can be predicted.

Fig. 6. Detection of double pores on the membrane with various probes. (a) Current traces obtained by scanning the probe with different tip diameters over two 20 nm-in-diameter nanopores. Their axis separation was set to 50 nm. The lateral distance represents the separation between the probe axis and the center of two nanopores. (b) Influence of a concentration gradient on the detection of two 20-nm-diameter pores with 50 nm in axis separation. The concentration gradient of 100: 300 mM KCl is set across the nanopores. (c-d) Detection of nanopores using a probe with equal inner diameter but different outer diameter at the tip. (c) Current traces over a 20-nm-diameter nanopore. (d) Detection of double pores with different diameters of 20 nm and 50 nm. The inner diameter of the probe is 60 nm. The outer probe diameter is considered as 110 and 260 nm, corresponding to 25 and 100 nm in thickness at the tip, respectively. The axis separation of both nanopores is 100 nm. The probe is located at 10 nm above the membrane.

Probe parameters are also considered in the detection of two nanopores on the membrane, including the inner and outer diameters at the tip. The considered outer probe diameter of 110 and 260 nm corresponds to 25 and 100 nm in thickness at the tip, respectively. From Fig. 5a, the glass probe with a tip diameter of 60 nm has limitations for the identification of the two nanopores with a small separation of 50 nm. In Fig. 6a, a probe with a smaller tip size of 30 nm is considered for the same detection. From the current profile, the small probe has a higher resolution for the location identification of nanopores. Based on the results of single pore detection under salt concentration gradients, the application of a salt concentration gradient across the porous membrane can also improve the detection current values (Fig. 6b), which can facilitate the detection of two nanopores located very close to each other.

During the approach of the glass nanopipette to the membrane, the thickness of the probe can modulate the confinement of the space between the nanopore and the probe. Here, the effect of the tip thickness of the probe on the detection of nanopores is explored using different probes with an equal inner diameter of 60 nm but different outer tip diameters. Two glass probes are selected with 110 and 260 nm in the outer tip diameter, corresponding to 25 and 100 nm in thickness at the pipette tip. From Fig. 6c, we find that the larger outer tip diameter can induce a decrease in the probe current when the probe is far away from the pore. This is caused by the higher confinement between the probe and the surface. As the probe scans over the nanopore, the probe current becomes similar to that using a thinner probe. The application of a probe with a larger outer diameter results in a larger current increase (Fig. 6d). This is good for the location detection of individual nanopores on the membrane. For multi-pore detection, the larger outer diameter can limit the detection accuracy due to its thicker membrane, which cannot provide the exact locations as shown in Fig. 6d. While the number of pores can also be detected clearly. In the case of two nanopores with an axis separation less than the thickness of the probe tip, to obtain the accurate location of the nanopores, two or more scans in different directions will be required to reduce the influence of the nearby nanopore.

## 4. Conclusions

COMSOL Multiphysics simulations have been conducted to mimic the detection of nanopores with the SICM technique. Results show that this technique can serve as an easy and fast *in situ* way for the identification of the nanopore location and number on porous membranes, due to the increase of the probe current by the transported ions through the nanopore. The detection accuracy can be affected by the selection of the tip diameter and thickness, as well as the separation between the probe and the membrane, which can be enhanced further by the application of a larger voltage or a concentration gradient across the nanopore. A larger glass probe is better for the detection of single nanopores. However, with multi-pores on the membrane, the detection sensibility is limited by the diameters of the glass probe, which cannot detect the pores with a small axial separation. For two nanopores located close to each other, glass probes with small diameters are preferred which have higher space resolution. In such a case with a thick wall at the probe tip, the accurate location of nanopores can be acquired through multi-scanning in different directions.

**CRediT authorship contribution statement**

**Yinghua Qiu**: Conceptualization, investigation, methodology, funding acquisition, writing – original draft, and writing – review & editing. **Long Ma**: Data curations. **Zhe Liu**: Data curations. **Hongwen Zhang**: Validation. **Bowen Ai**: Formal analysis. **Xinman Tu**: Supervision.**Declaration of competing interest**

The authors declare that they have no known competing financial interests or personal relationships that could have appeared to influence the work reported in this paper.

**Data availability**

Data will be made available on request.

**Acknowledgements**

This research was supported by the National Science Foundation of China (52105579), the Guangdong Basic and Applied Basic Research Foundation (2023A1515012931), the Open Foundation of Key Laboratory of Jiangxi Province for Persistent Pollutants Control and Resources Recycle, Nanchang Hangkong University (ES202380052), and the Qilu Talented Young Scholar Program of Shandong University.

**Appendix A. Supplementary material**

Supplementary data to this article can be found online.

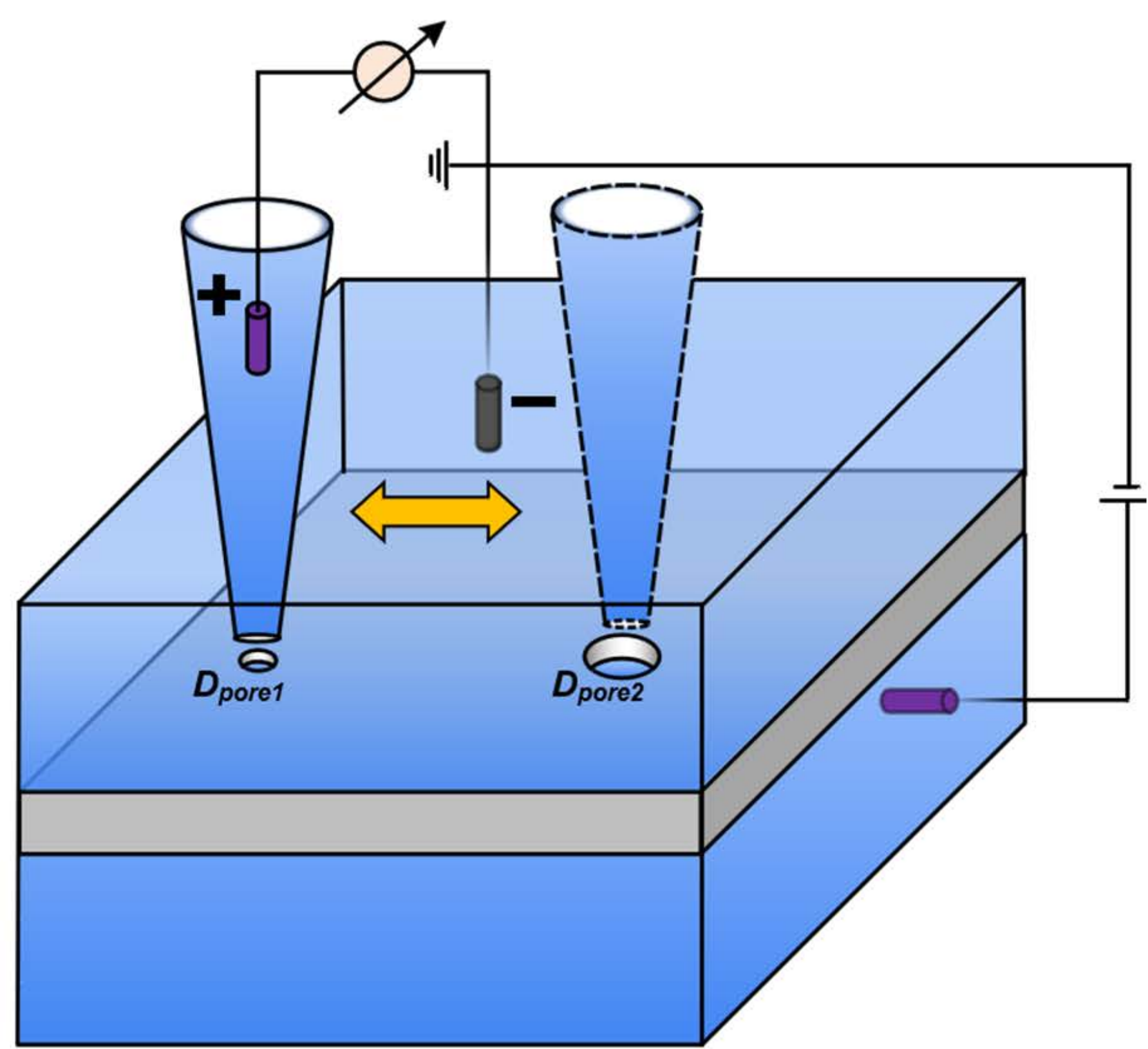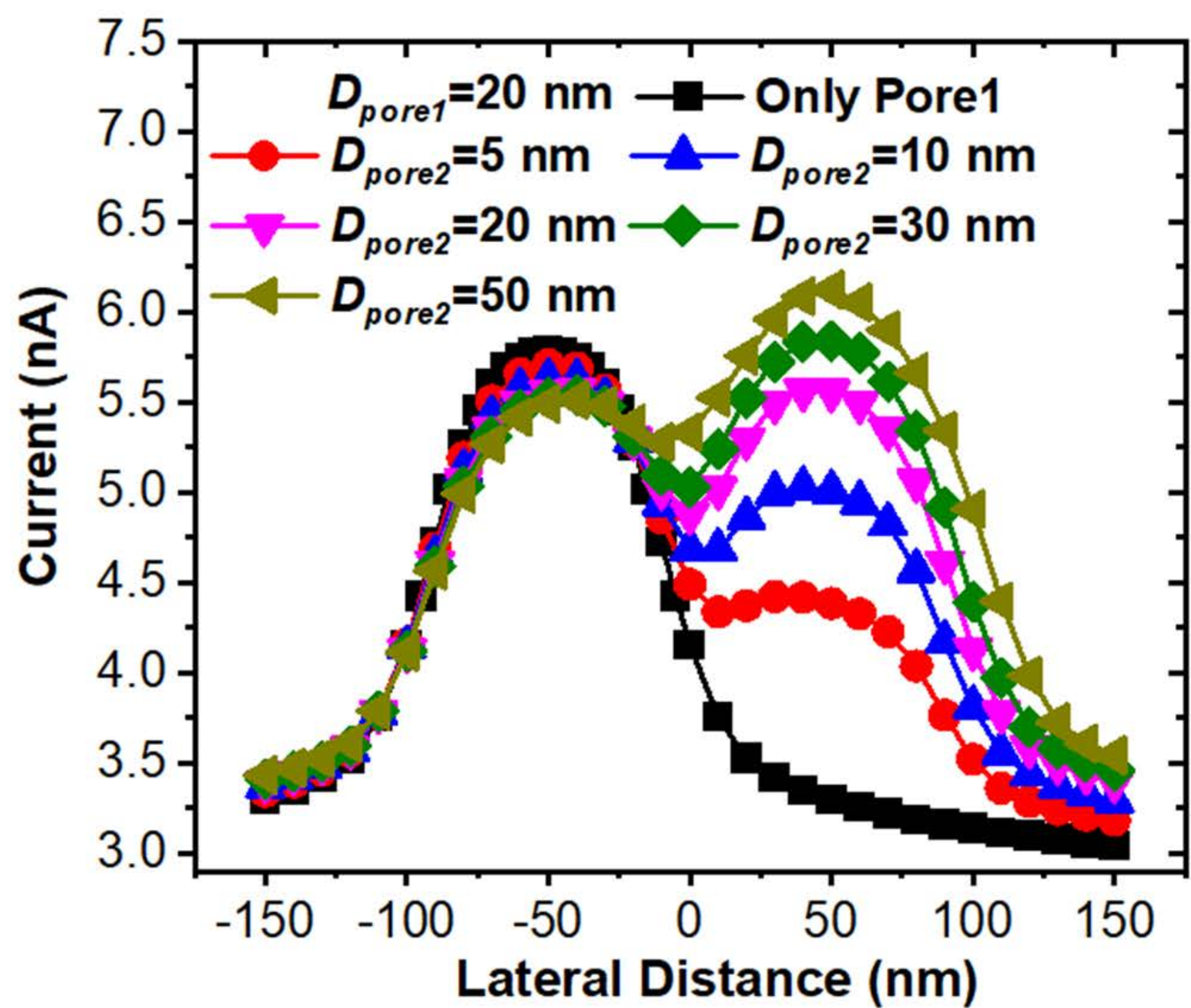

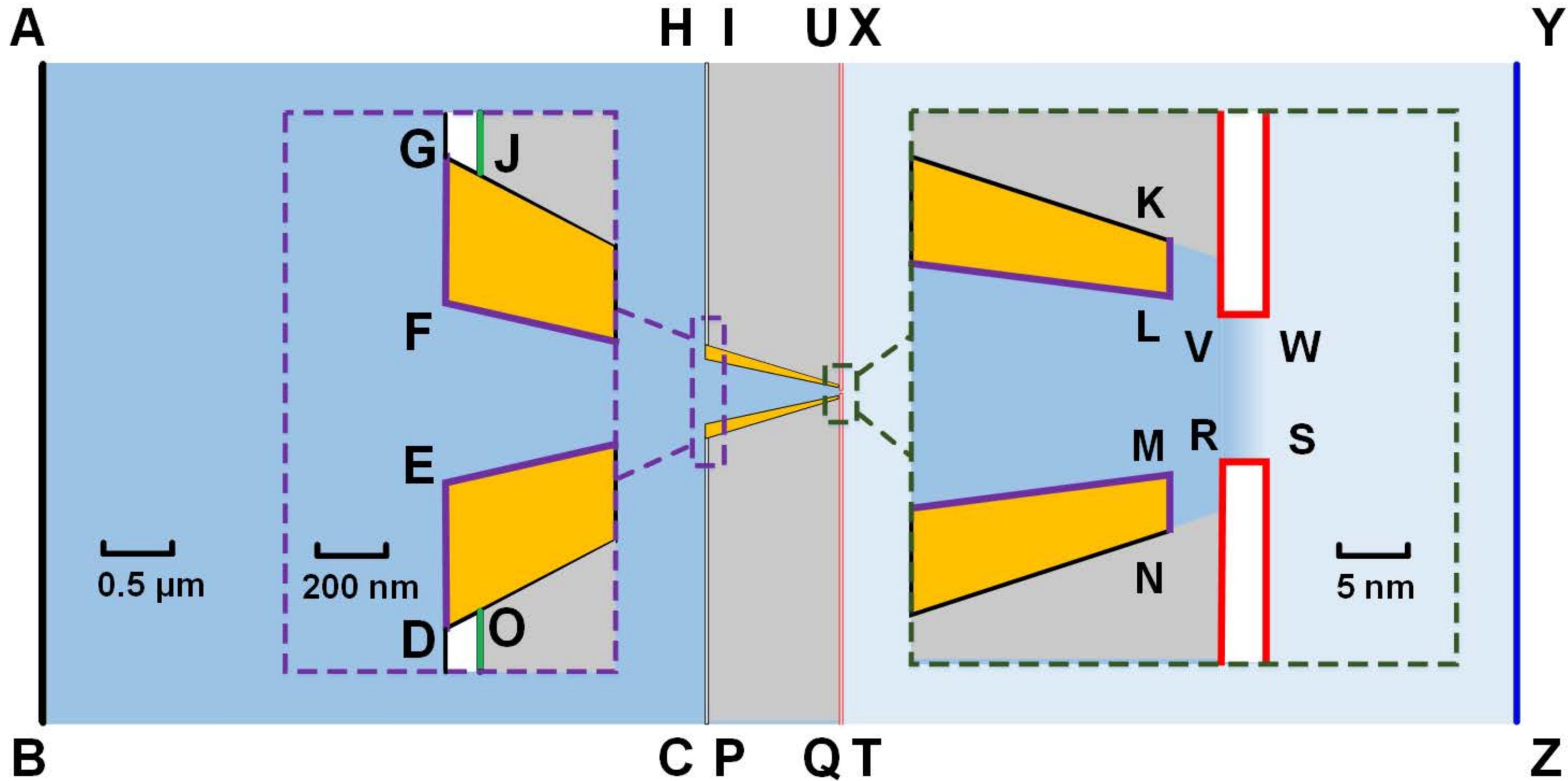

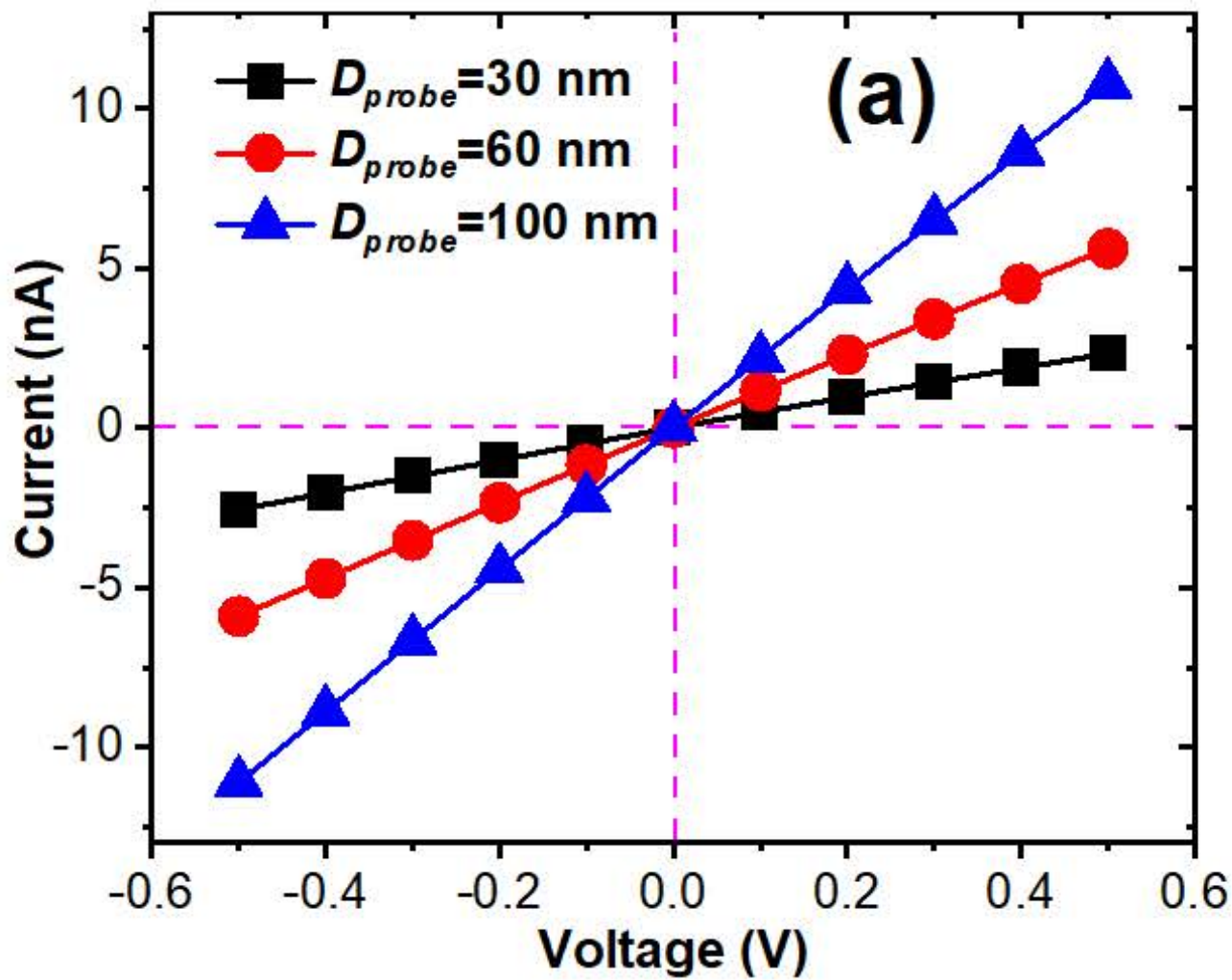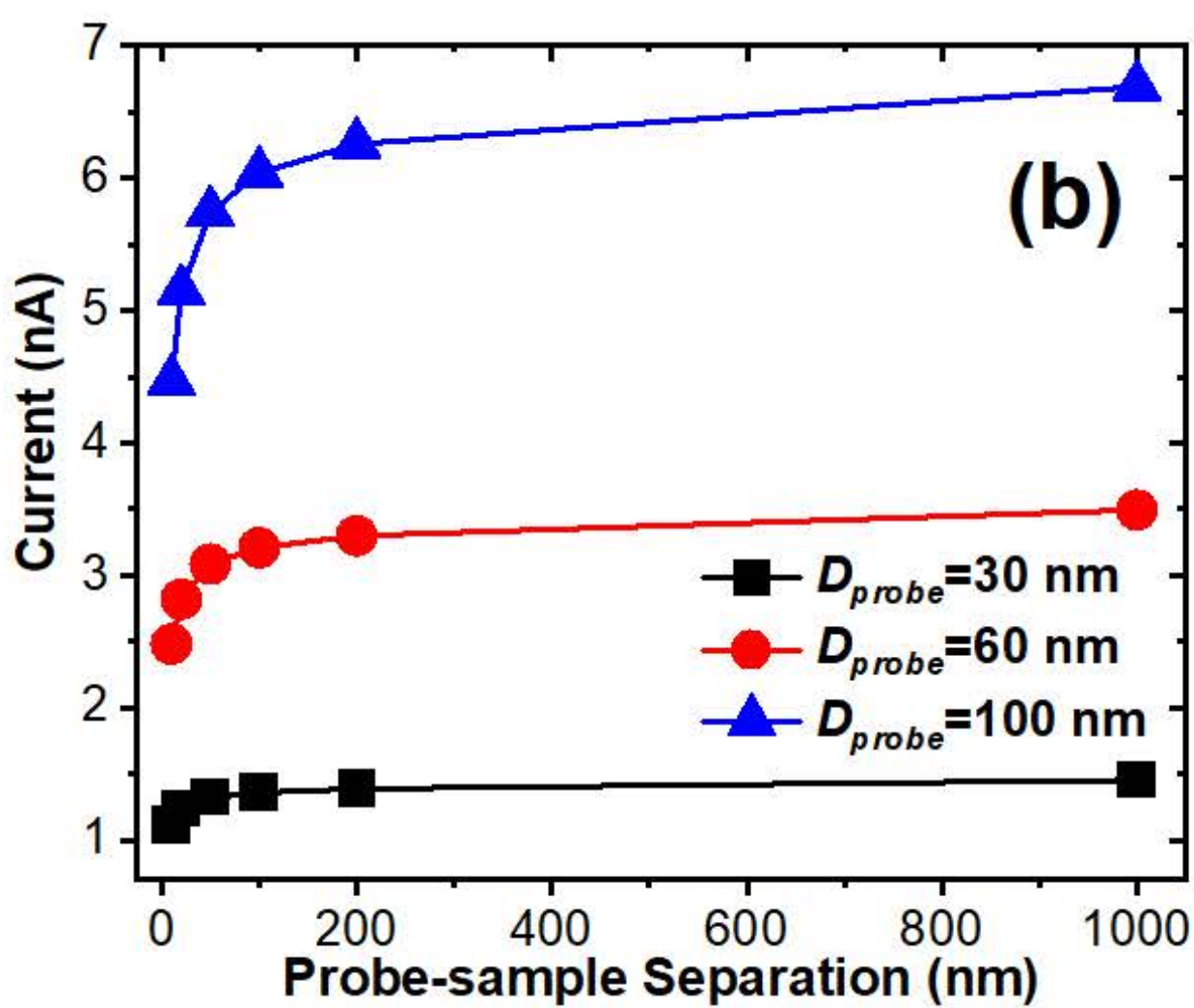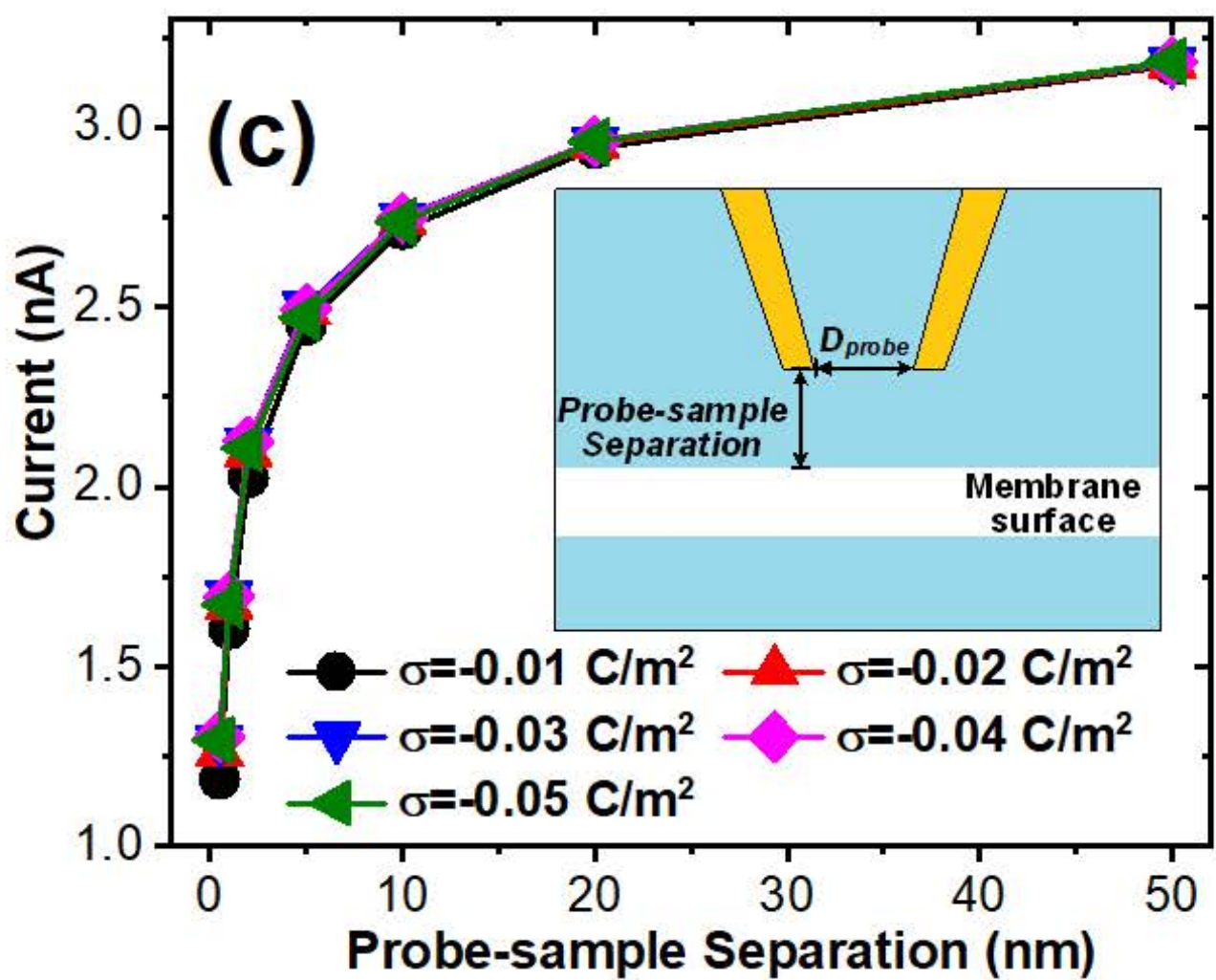

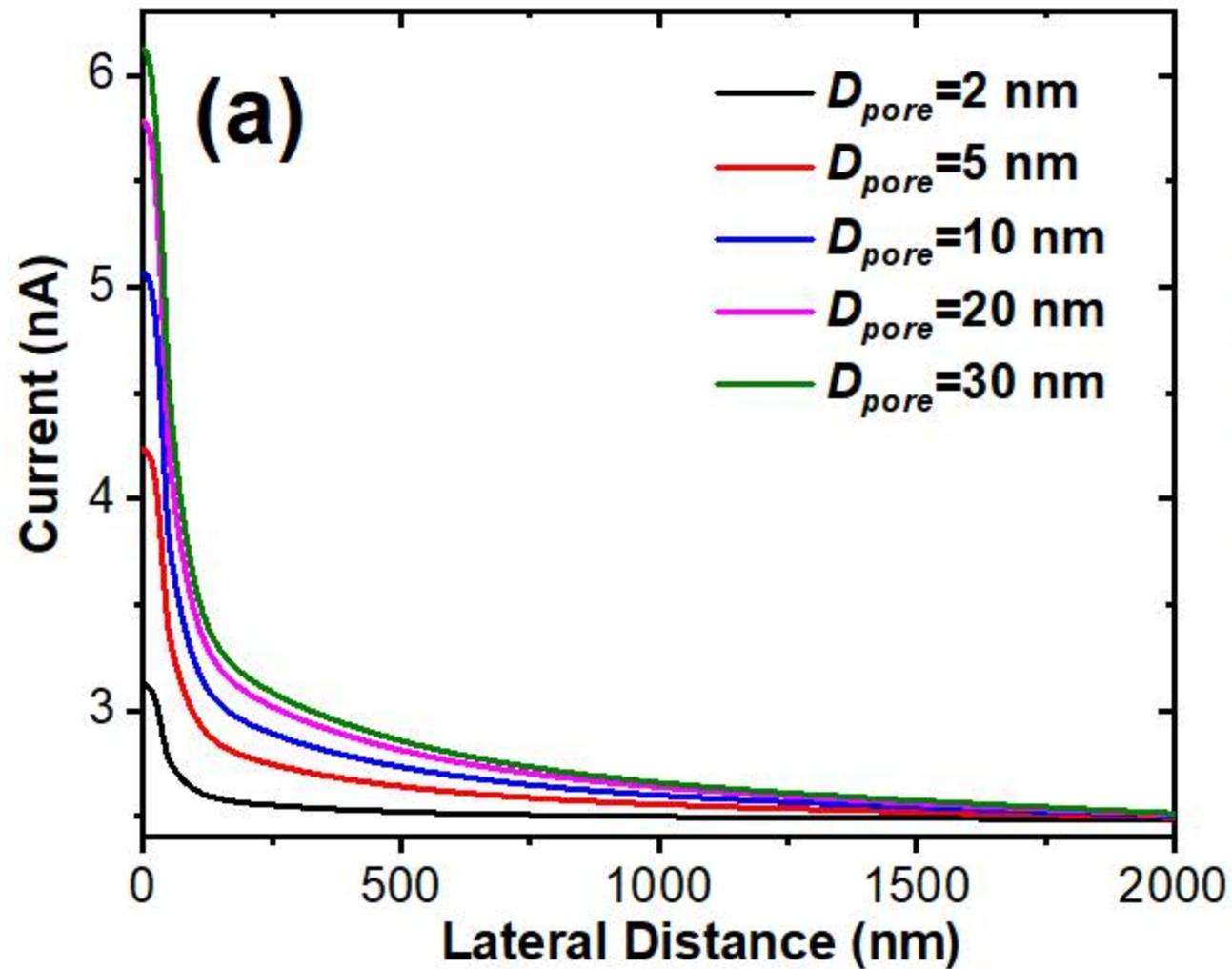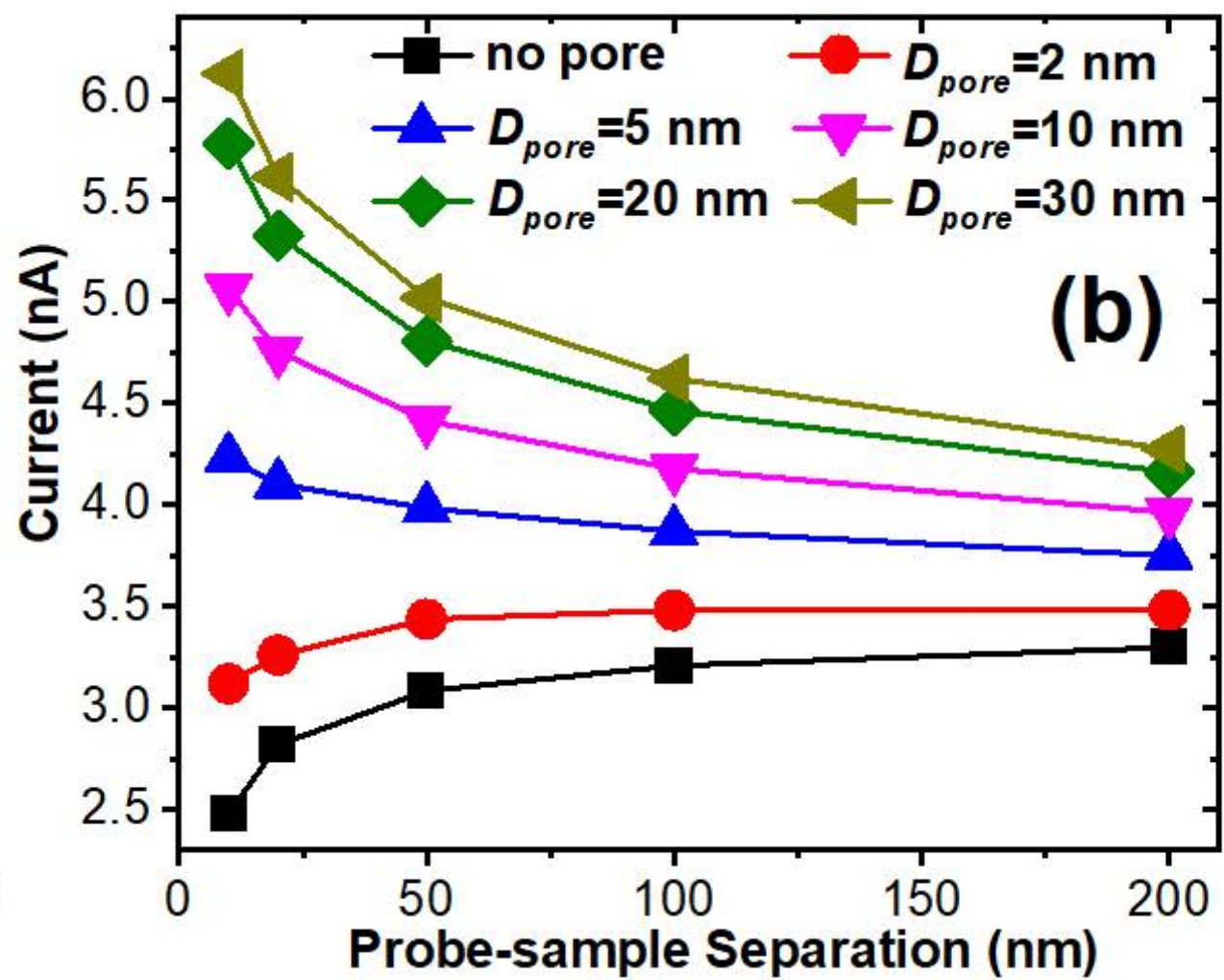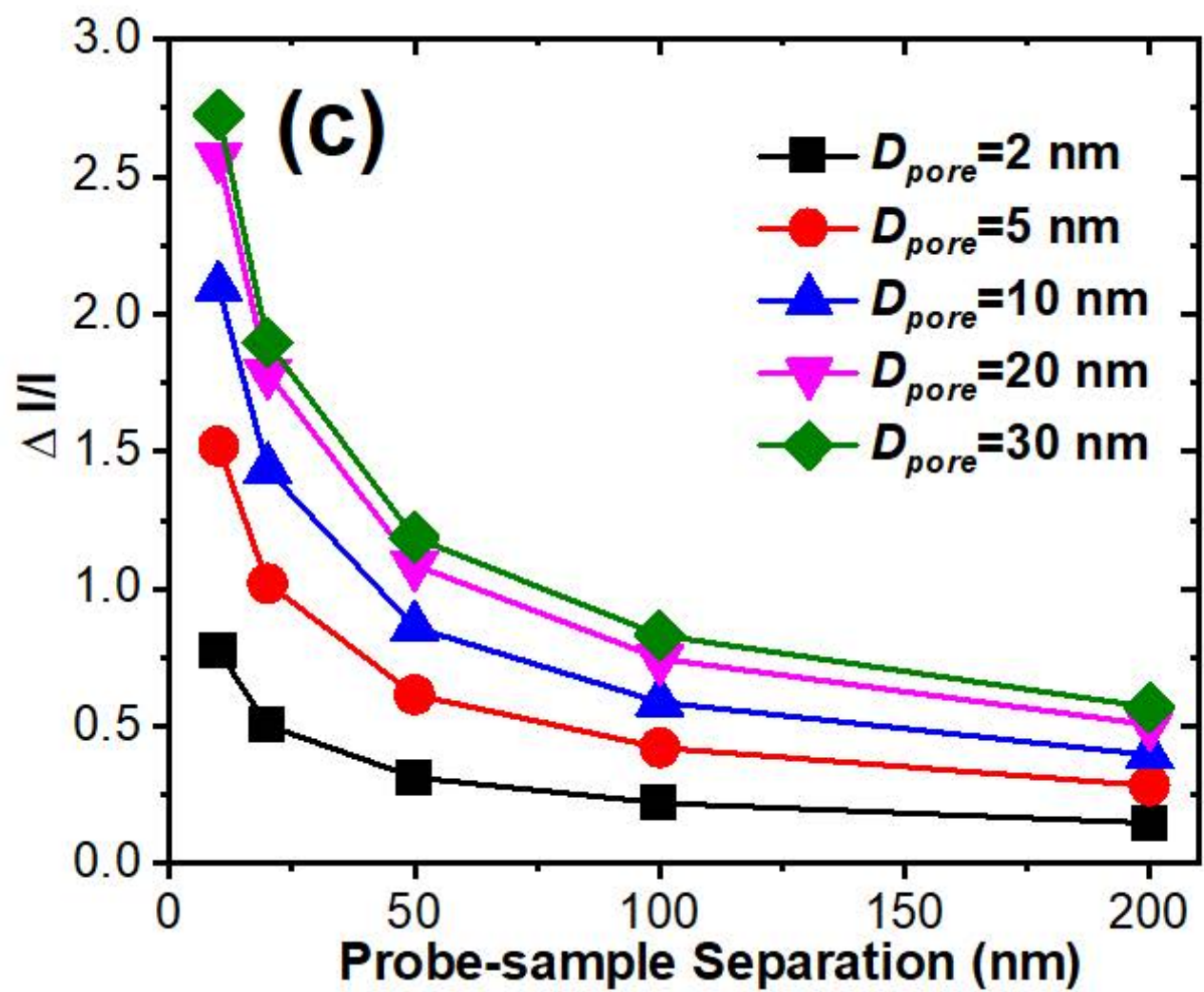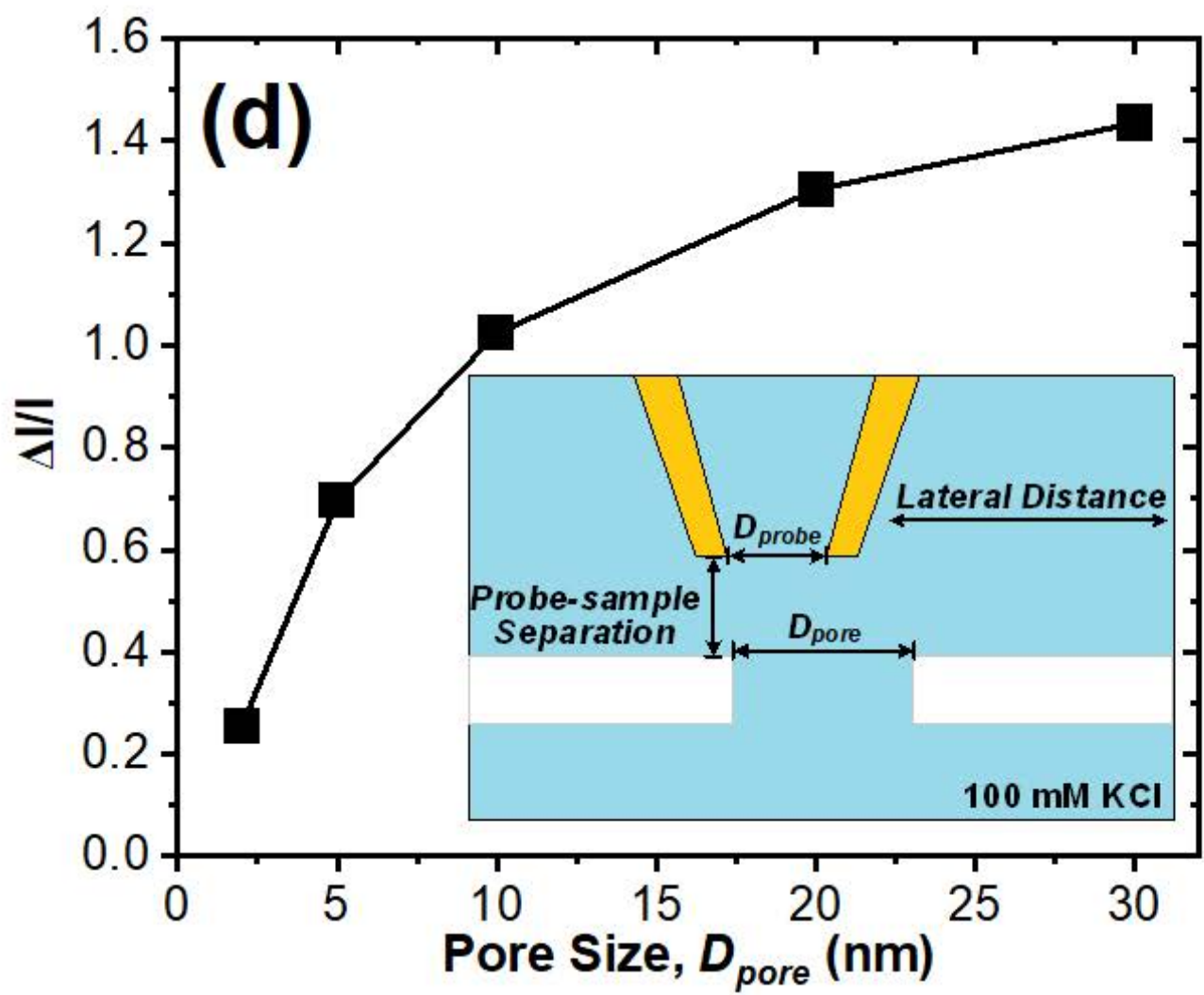

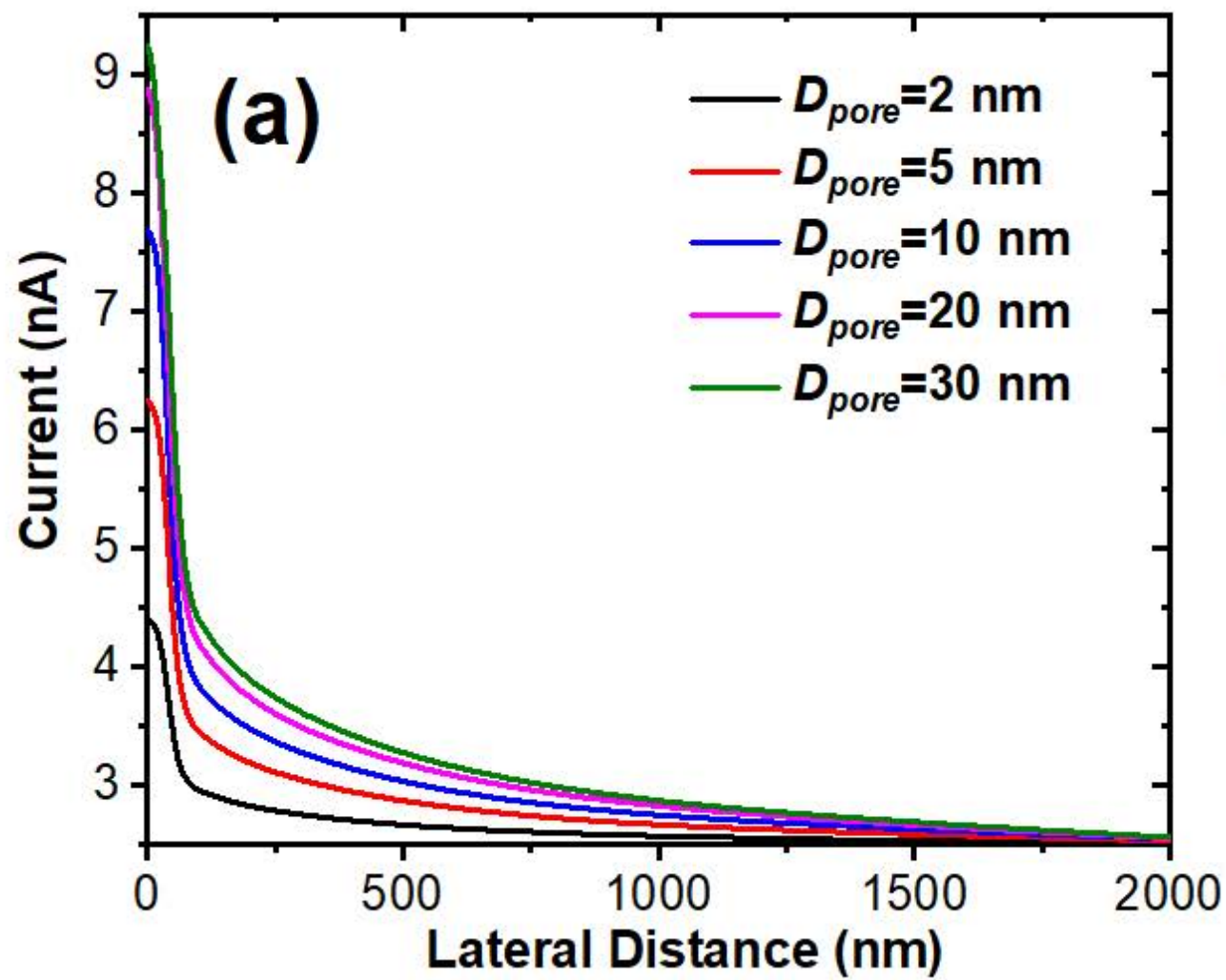
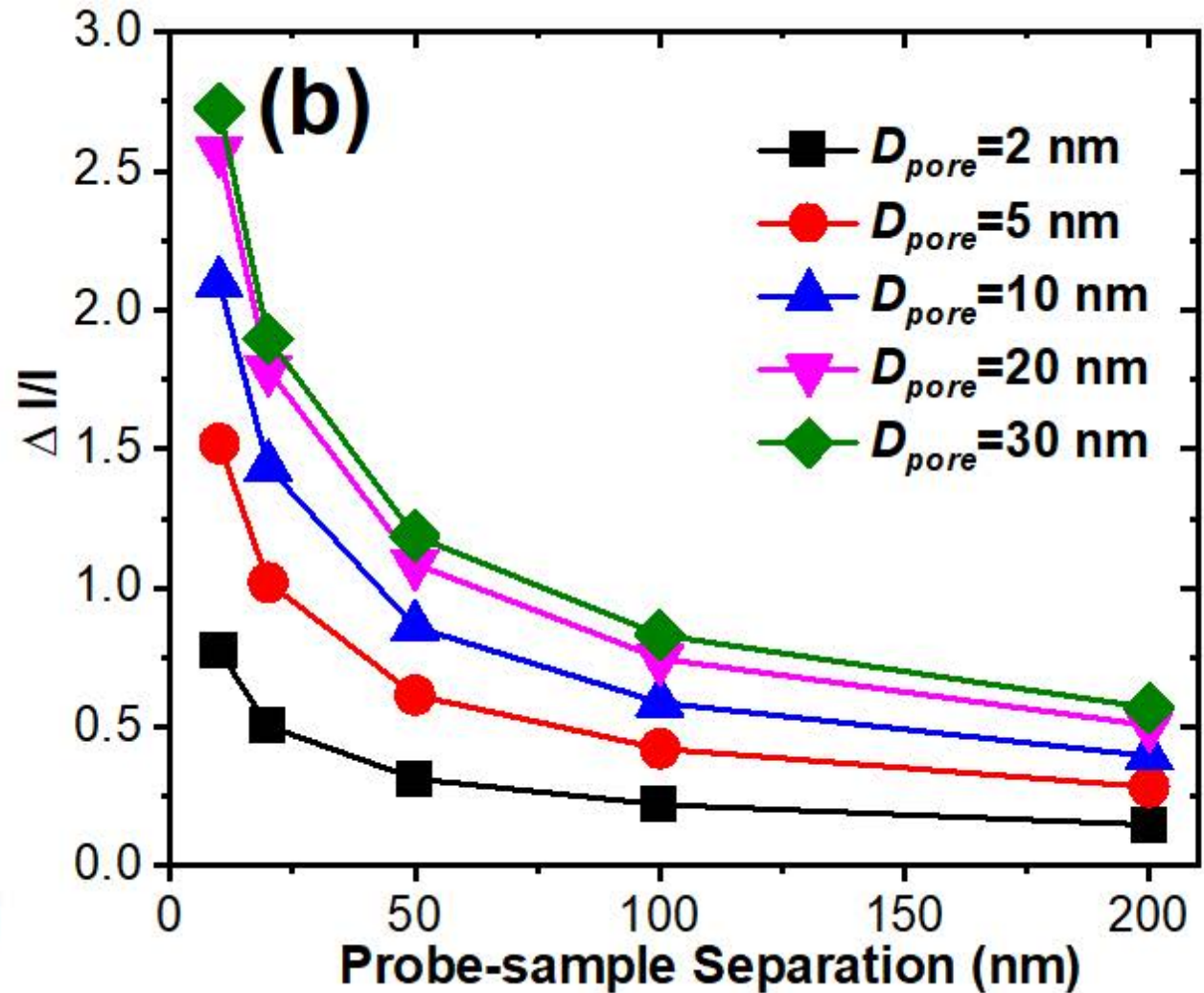
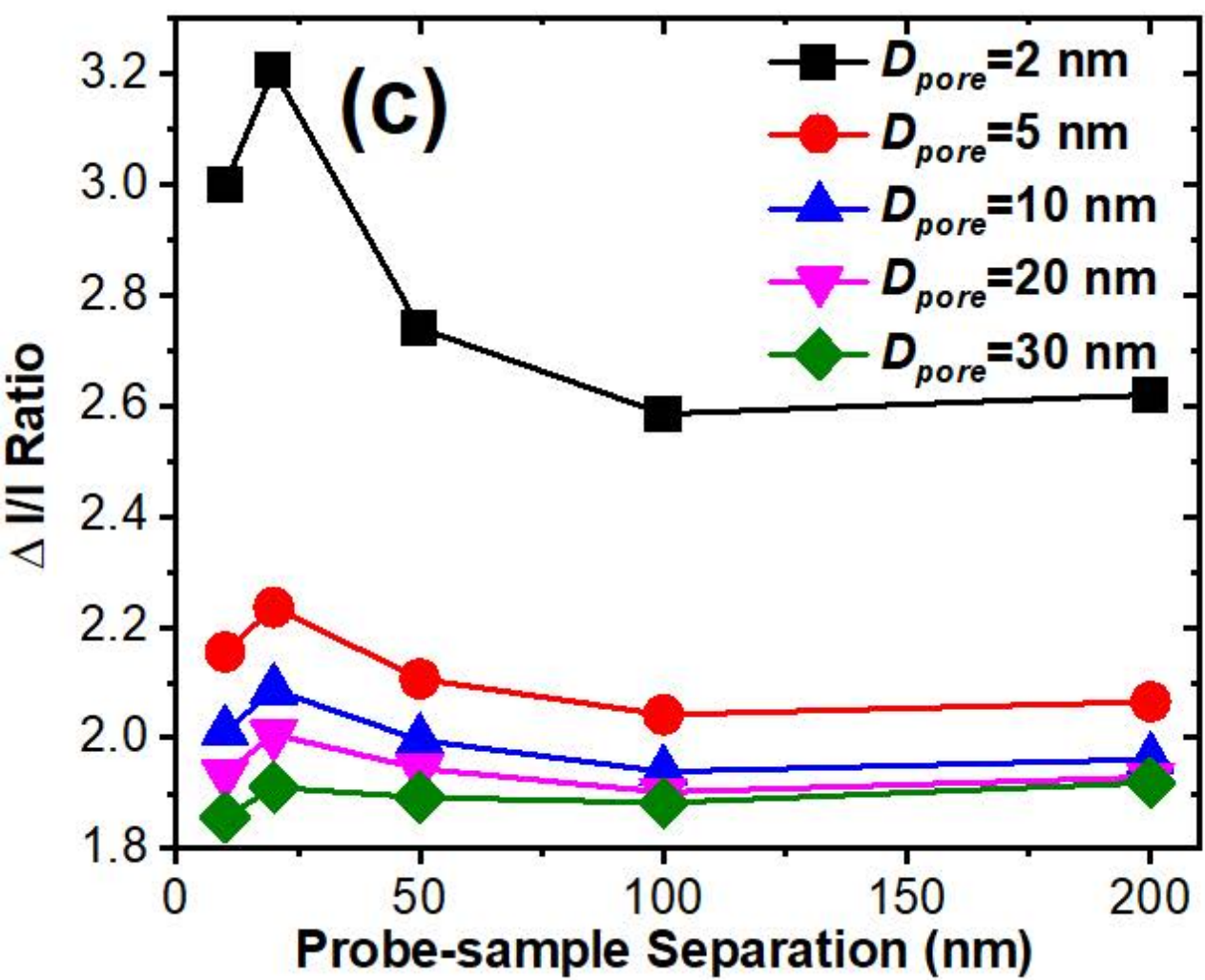
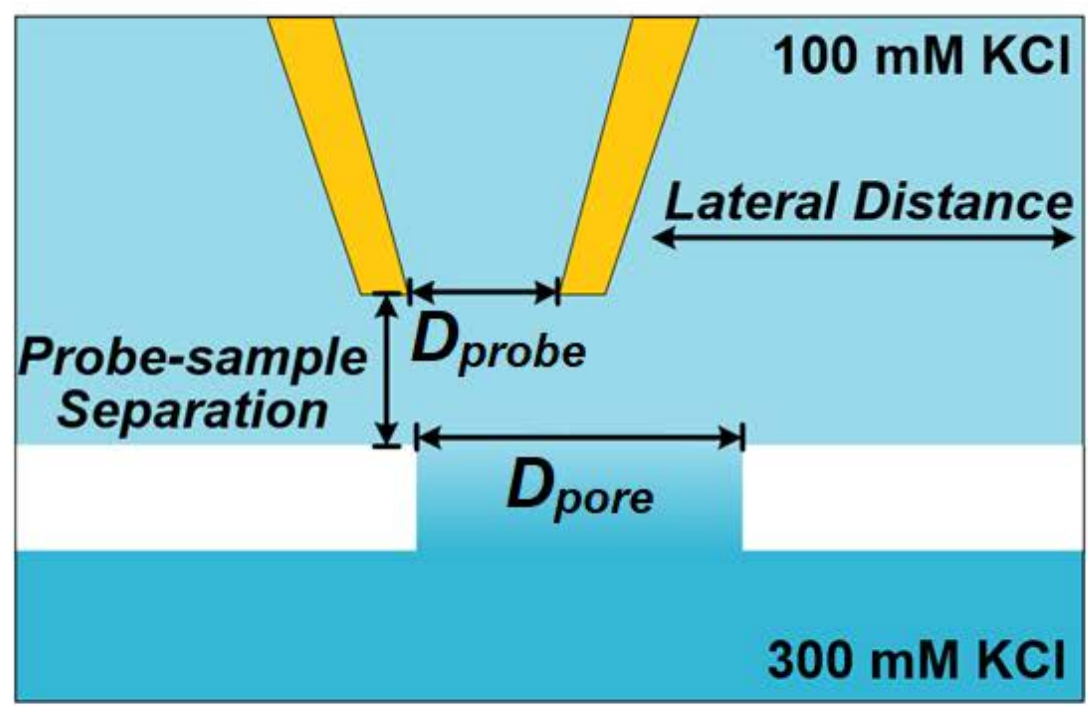

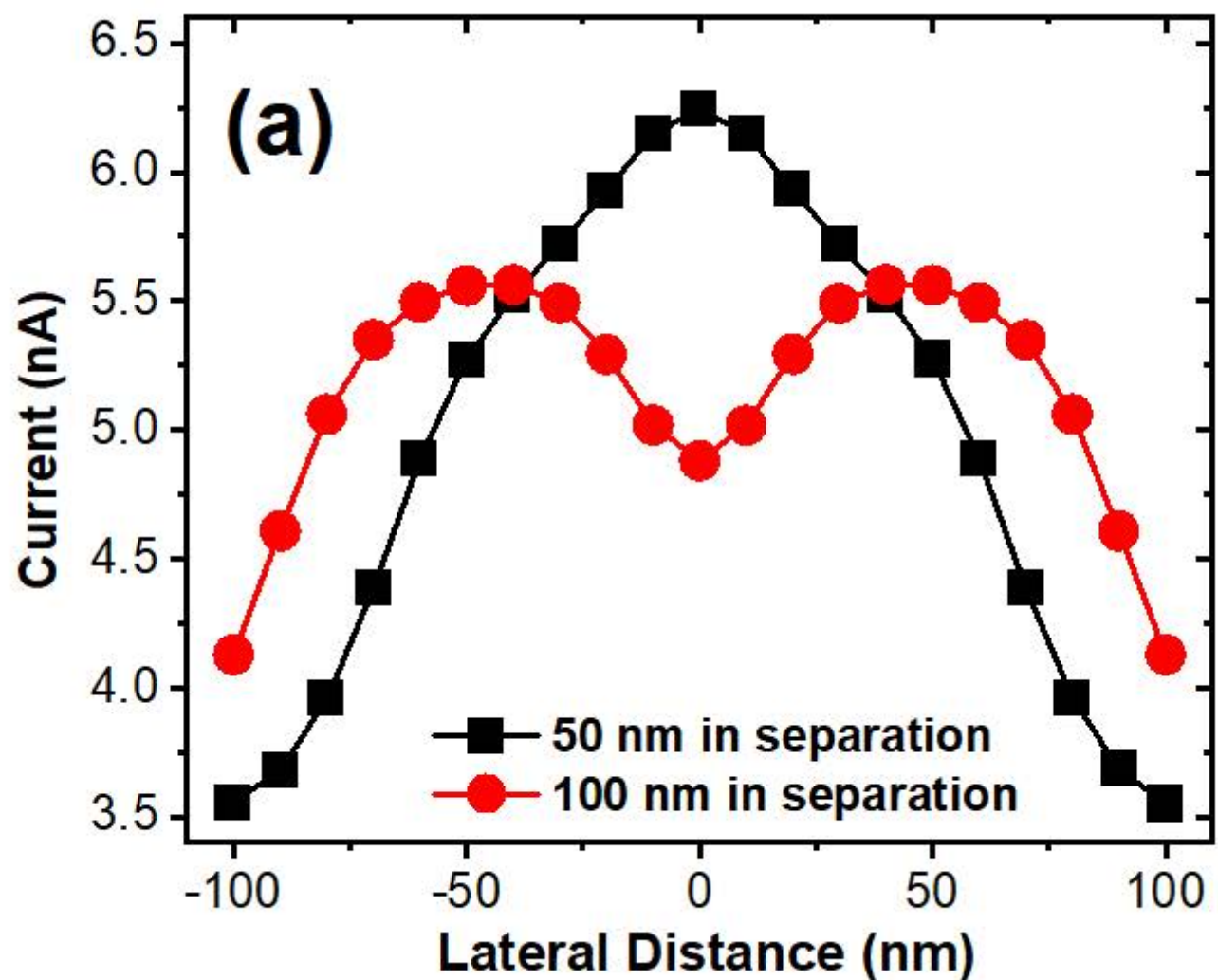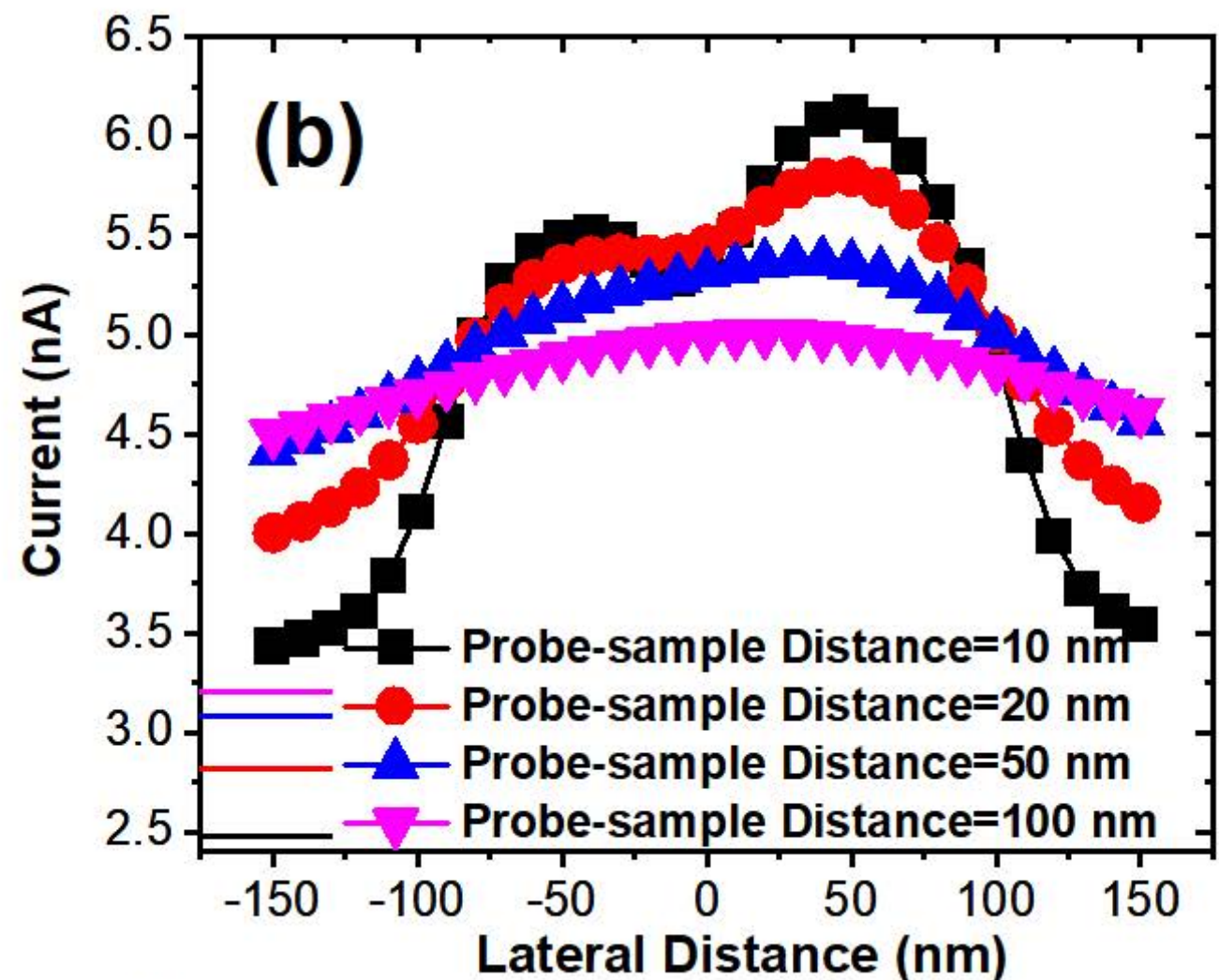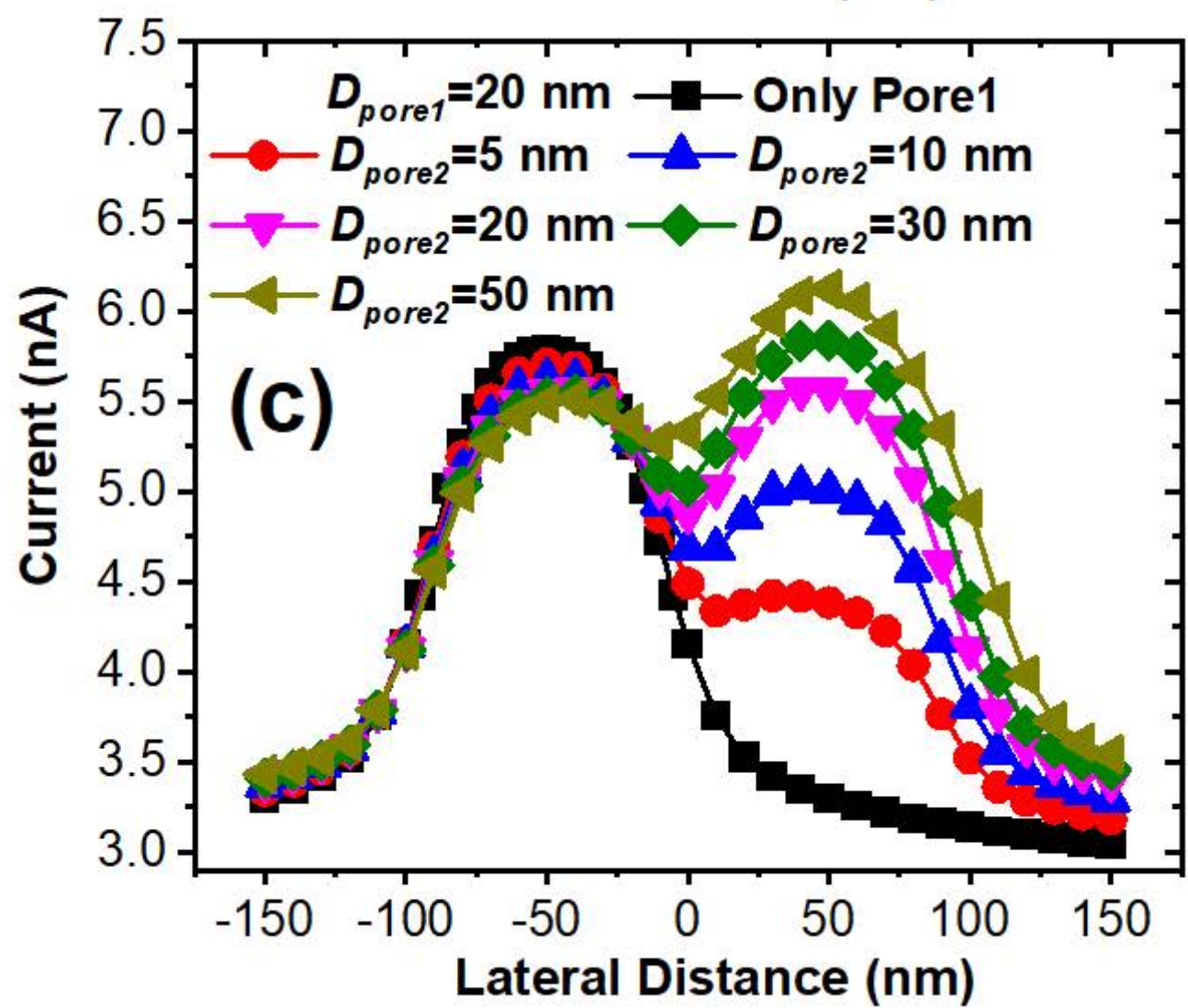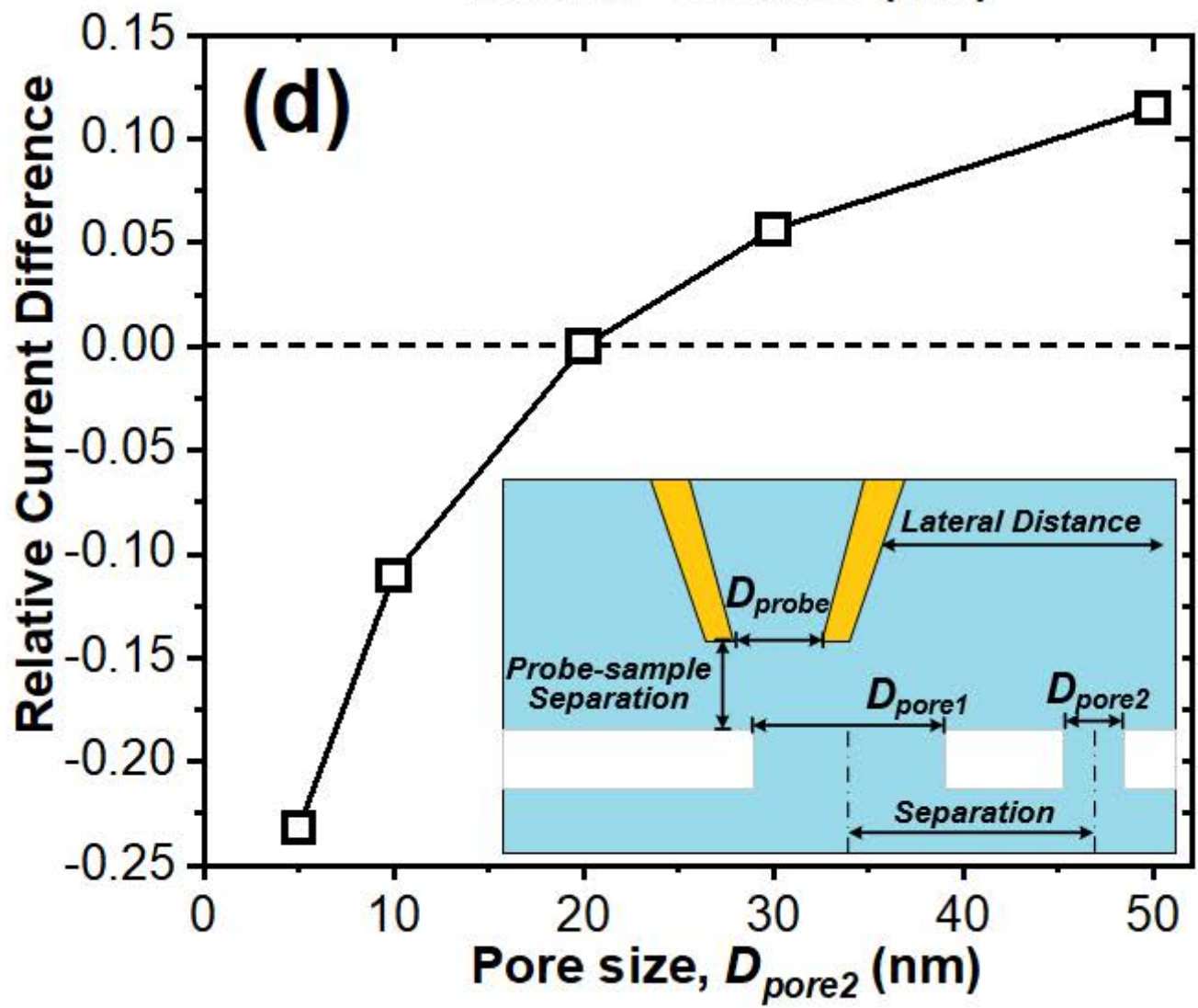

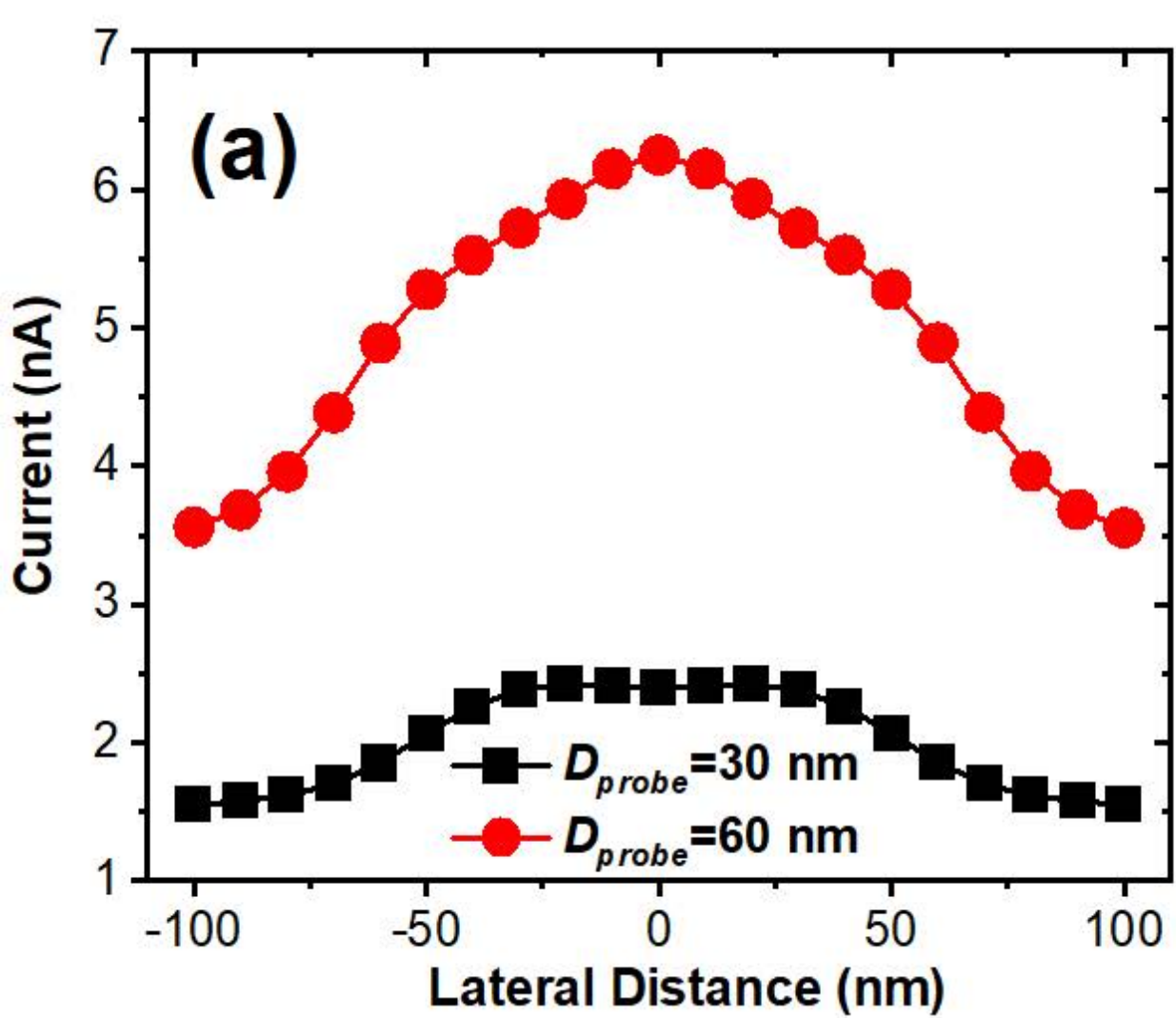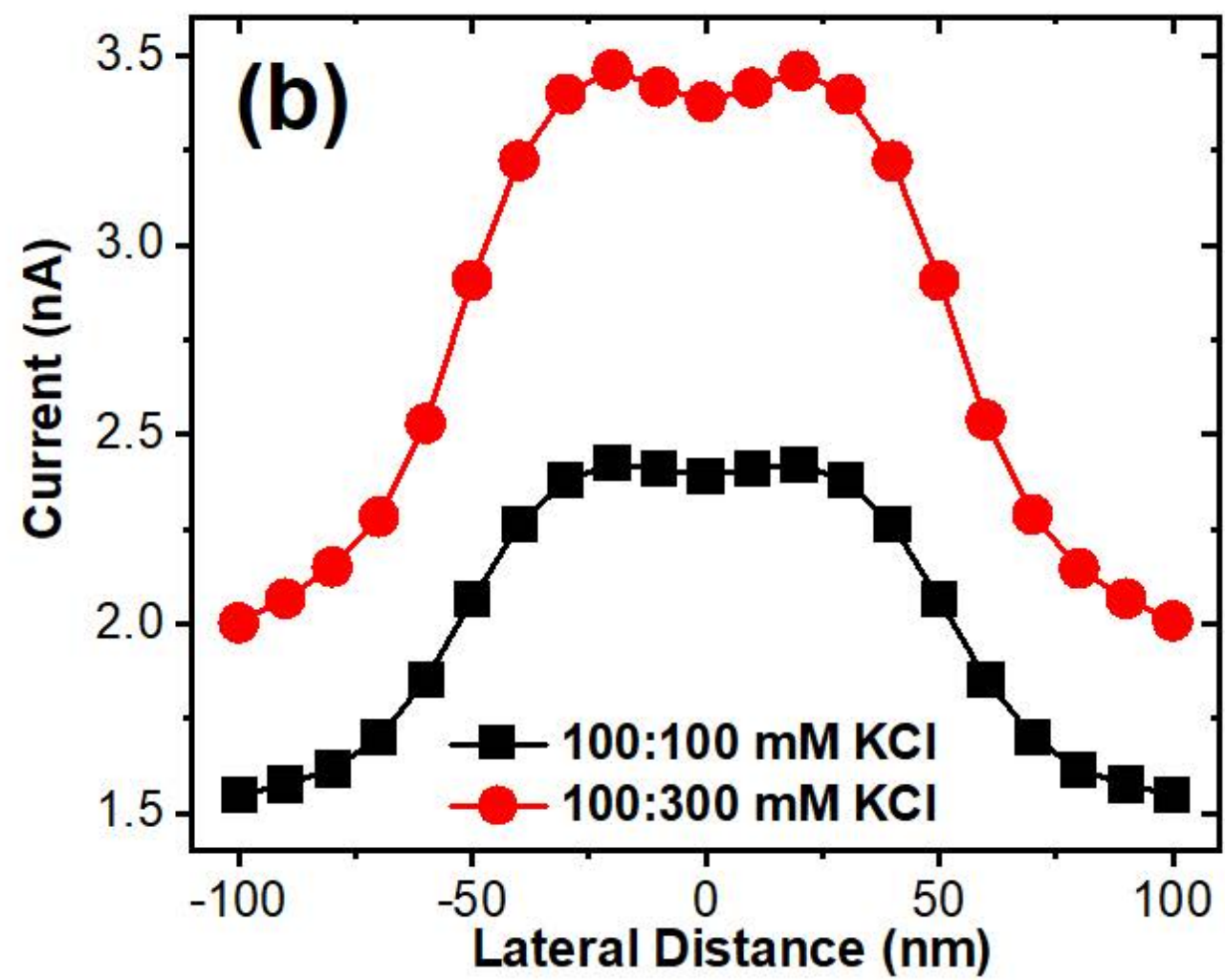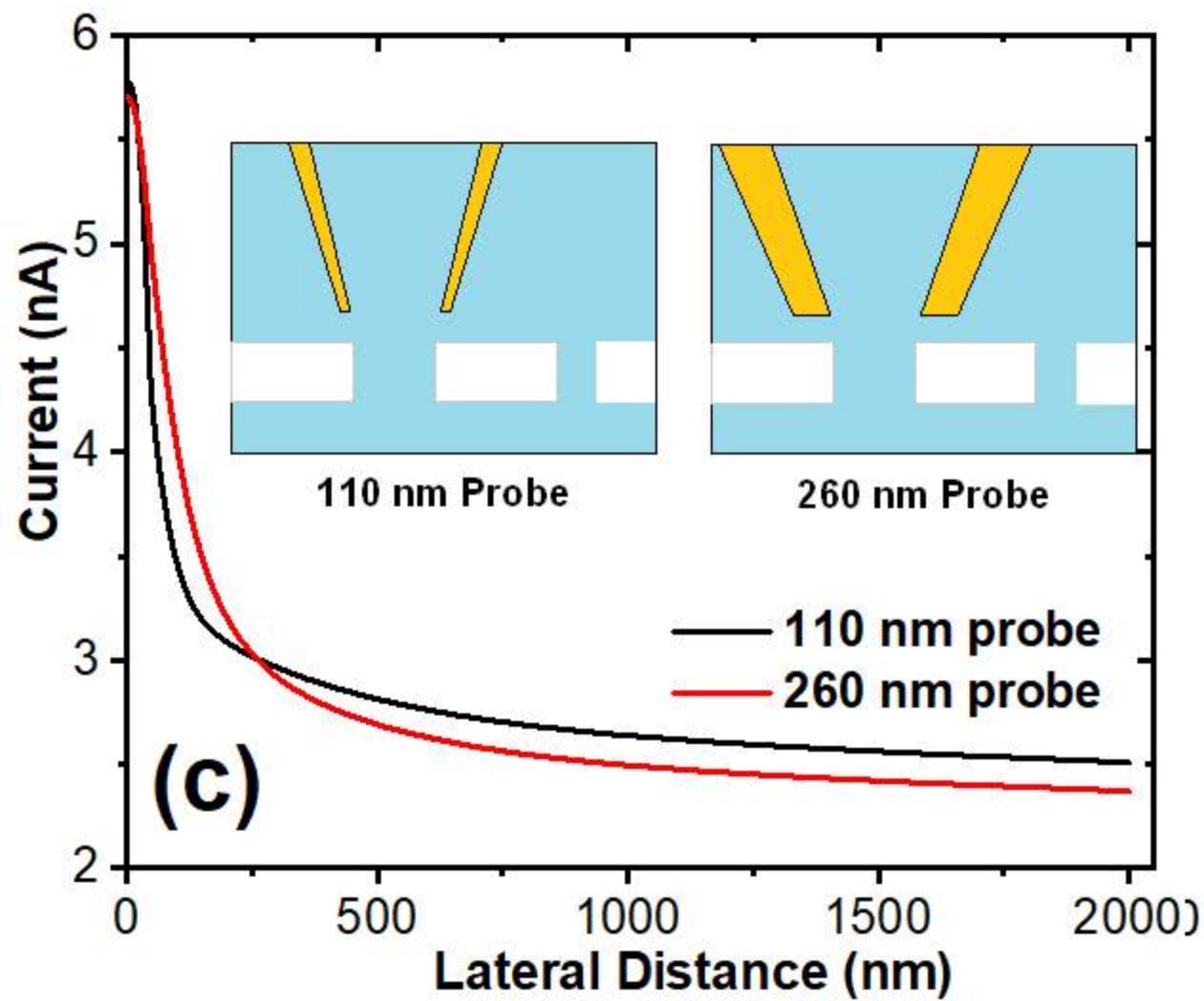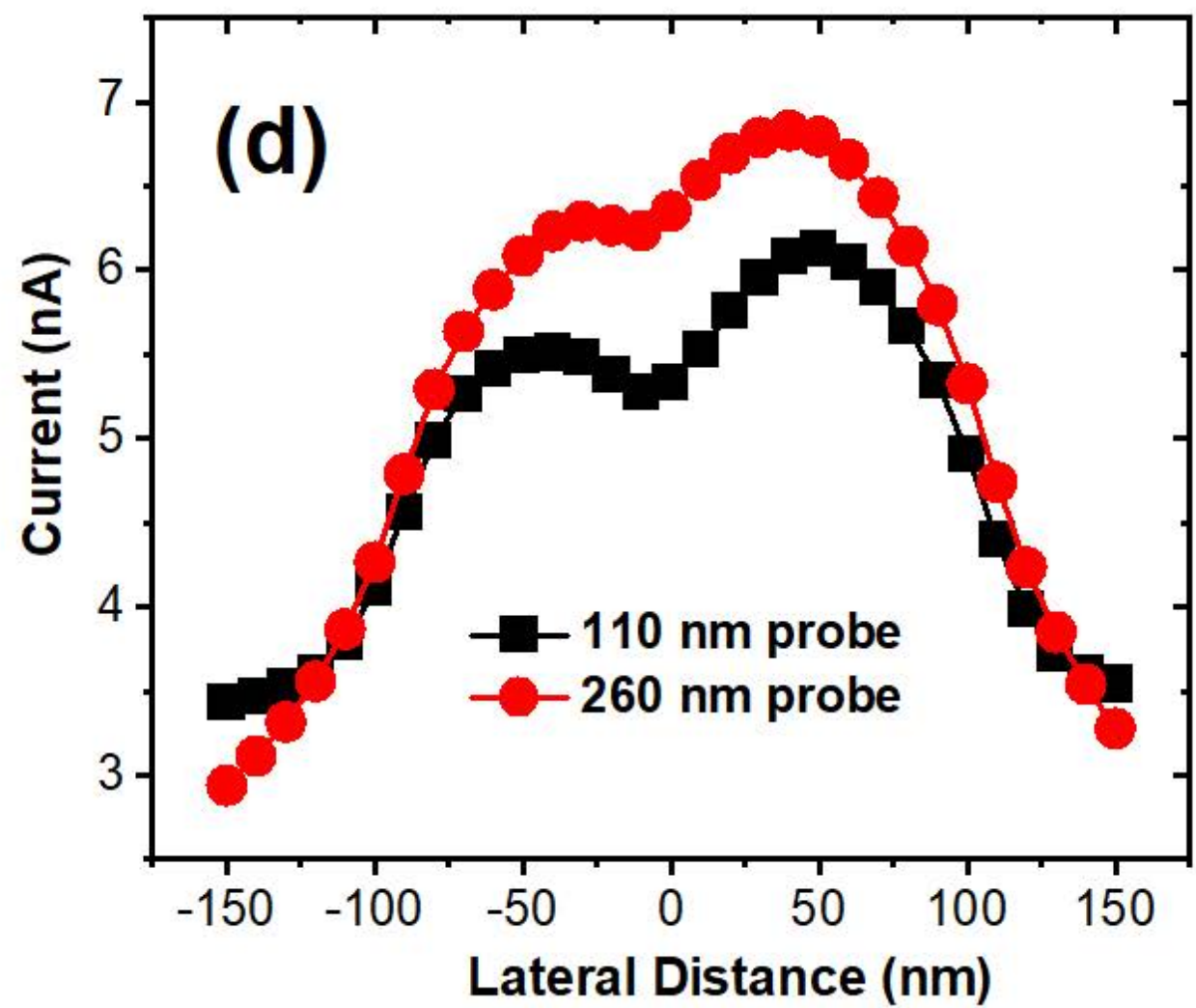

*Supplementary material*

# Detection of Nanopores with the Scanning Ion Conductance Microscopy: A Simulation Study

Yinghua Qiu,[1,2,3]* Long Ma,[2,3] Zhe Liu,[2,3] Hongwen Zhang,[2,3] Bowen Ai,[2,3] and Xinman Tu[1*]

1. Key Laboratory of Jiangxi Province for Persistent Pollutants Control and Resources Recycle, Nanchang Hangkong University, Nanchang, Jiangxi, 330063, China

2. Key Laboratory of High Efficiency and Clean Mechanical Manufacture of Ministry of Education, National Demonstration Center for Experimental Mechanical Engineering Education, School of Mechanical Engineering, Shandong University, Jinan, 250061, China

3. Shenzhen Research Institute of Shandong University, Shenzhen, Guangdong, 518000, China

*Corresponding author:  yinghua.qiu@sdu.edu.cn ; tuxinman@126.com

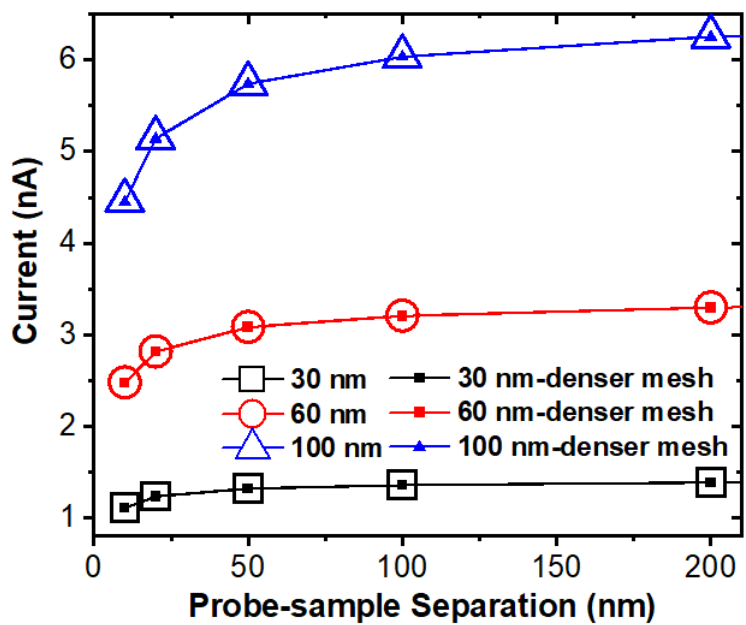

Figure S1 Independence of the results on selected mesh size. Larger symbols are current obtained with the mesh described in the manuscript. The denser mesh strategy is: For the probe, a mesh size of 0.5 nm was used for the charged surfaces (GF, FL, KL, DE, EM, MN). For the charged boundary of the reservoirs (UV, WX, QR, ST) the mesh of 1.25 nm was chosen. A finer mesh of 0.1 nm was selected for the inner surface of the nanopores.

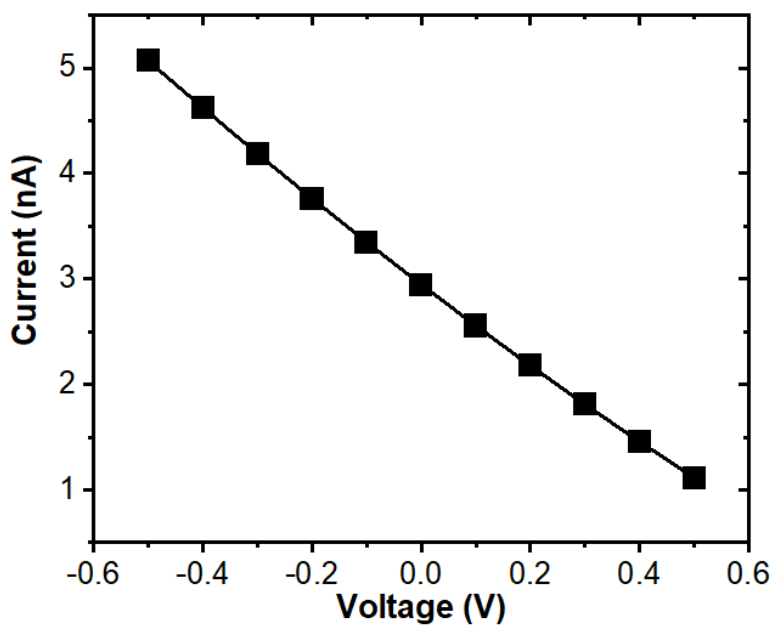

Figure S2 Dependence of probe current on the voltage applied on the nanopore side. The probe was 1 μm in length, 60 nm, and 500 nm in tip and base diameter, respectively. 0.3 V was used for the probe voltage.

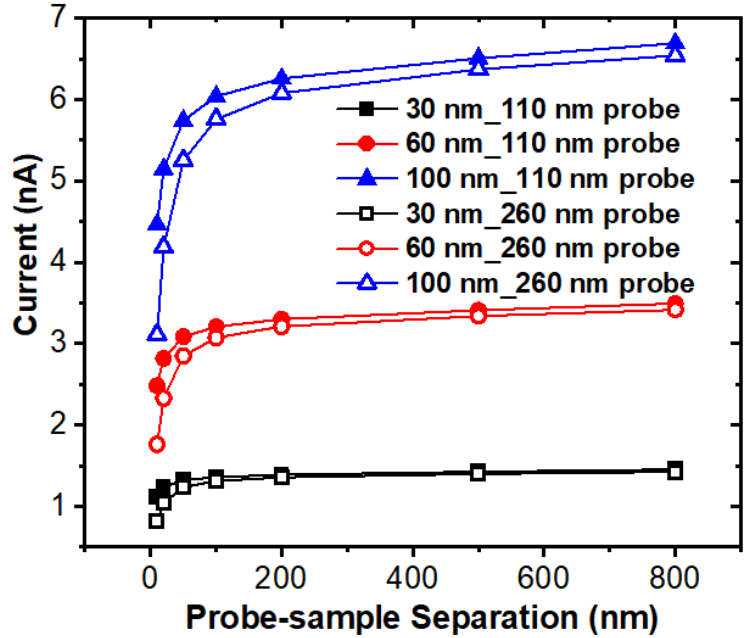

Figure S3 Currents of the glass probe approaching a surface location in front of the probe under 0.3 V.

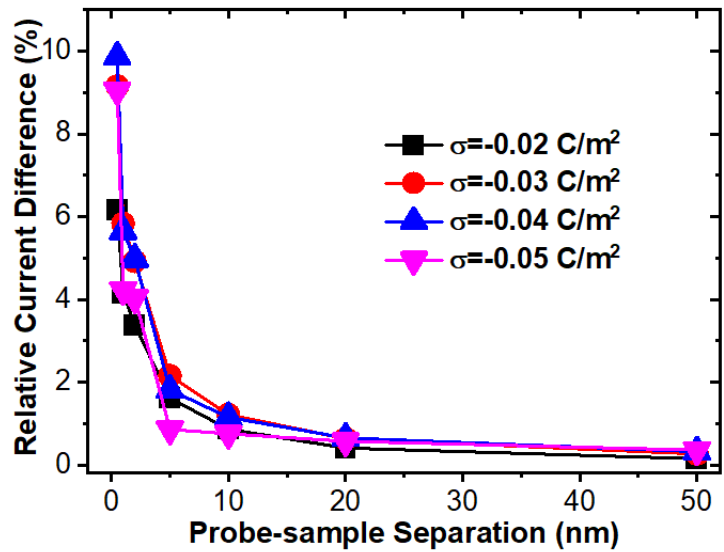

Figure S4 Relative current difference when the probe approaching the membrane with different surface charge densities compared with the values when the probe approaching the membrane with -0.01 C/m$^2$.

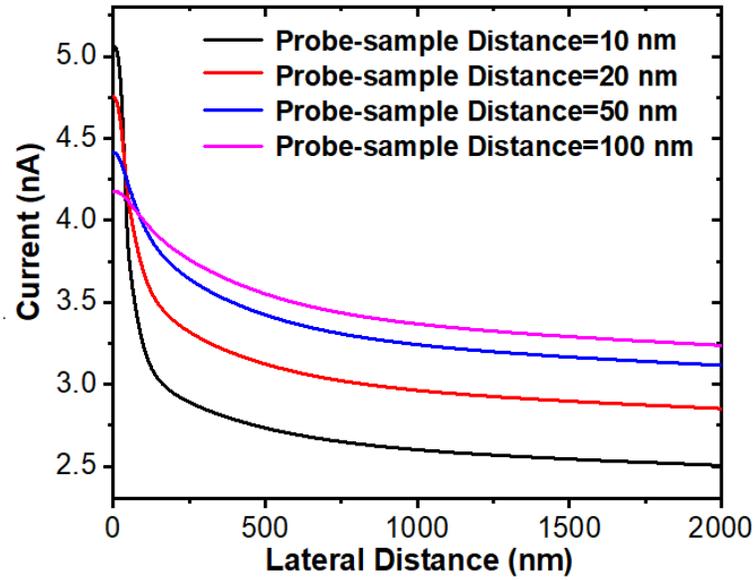

Figure S5 Current detection by a probe scanning in the radial direction of a 10 nm nanopore with separation above the surface.

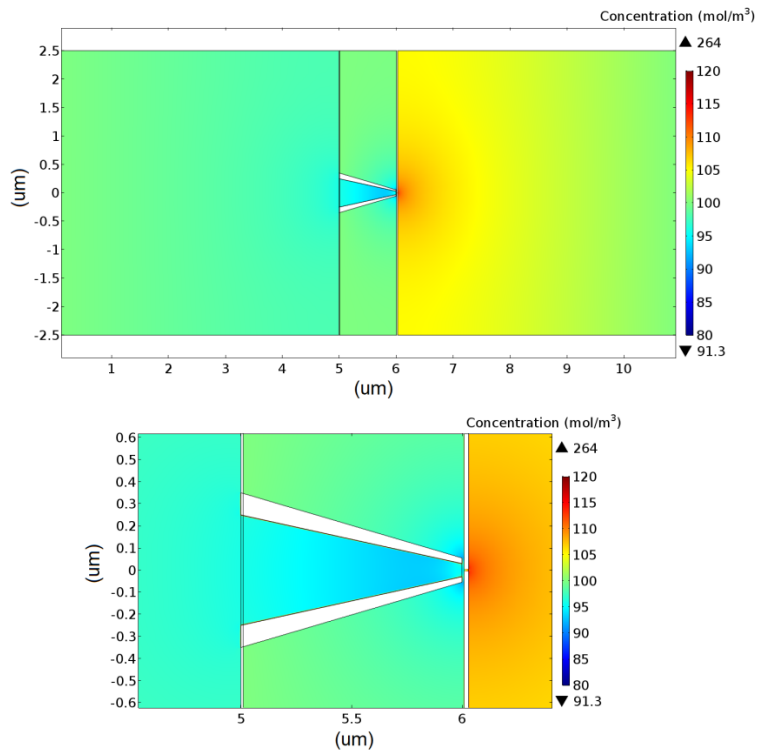

Figure S6 Concentration distribution of $K^+$ ions across the nanopore with 10 nm in diameter under −0.5 V. The bulk concentration is 100 mM.

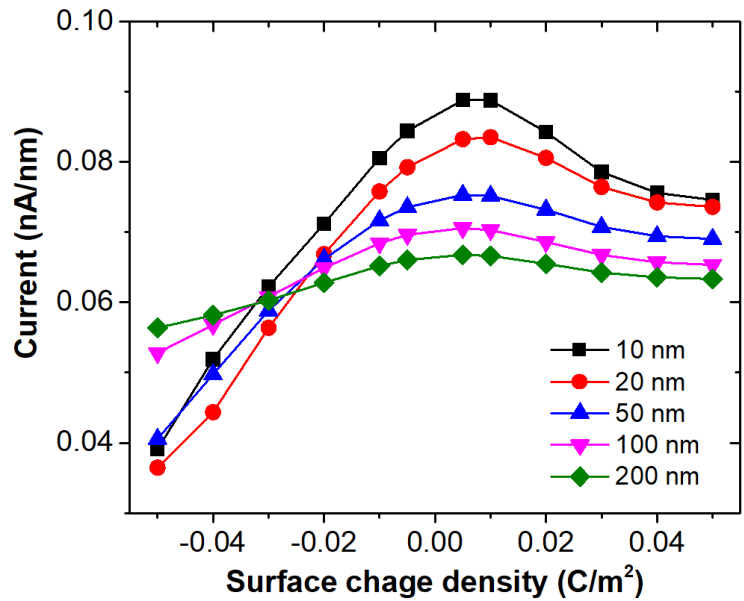

Figure S7 Dependence of probe current with the surface charge density of nanopores with 10 nm separation between the probe and membrane.

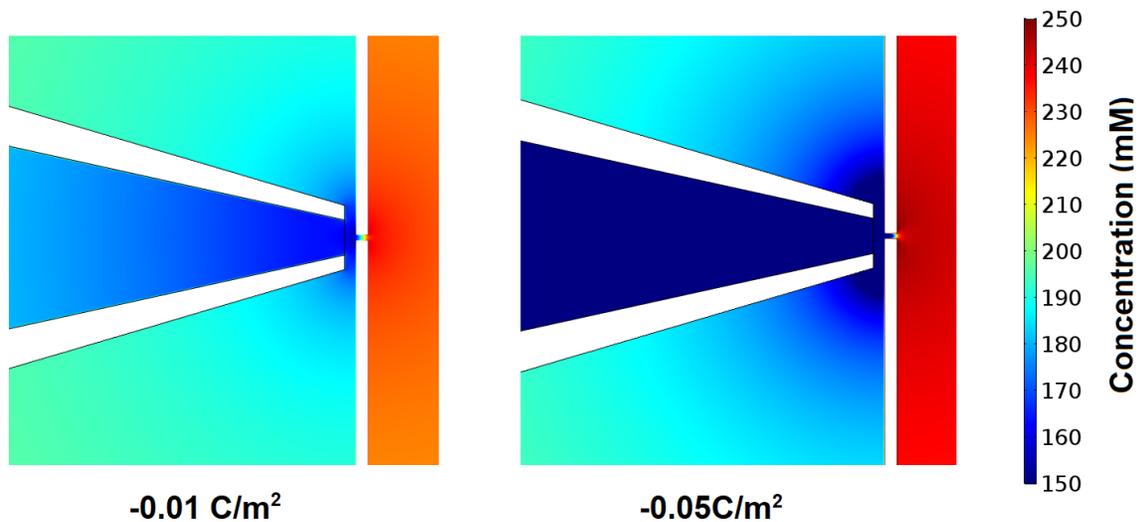

-0.01 C/m²        -0.05 C/m²

Figure S8 Concentration distribution of K$^+$ plus Cl$^-$ ions across the nanopore charged as −0.01 and −0.05 C/m² with 10 nm in diameter under −0.5 V. The bulk concentration is 100 mM. The separation was 20 nm.